\documentstyle[aps, preprint, tighten, eqsecnum]{revtex}
\begin{document}

\title{EMERGENT CHIRAL SYMMETRY: PARITY AND TIME REVERSAL DOUBLES}
\author{A. P. Balachandran 
         and S. Vaidya \footnote{sachin@suhep.phy.syr.edu}
}
\address{\em Department of Physics, Syracuse University,
             \\ Syracuse, N . Y.  13244-1130,  U. S. A. 
        \\
}
\preprint{\vbox{\hbox{SU-4240-653} \hbox{February 1997}}}

\maketitle

\begin{abstract}
There are numerous examples of approximately degenerate states of opposite
parity in molecular physics. Theory indicates that these doubles can occur
in molecules that are reflection-asymmetric. Such parity doubles occur in
nuclear physics as well, among nuclei with odd A $\sim$ 219-229. We have
also suggested elsewhere that such doubles occur in particle physics for
baryons made up of {\it cbu} and {\it cbd} quarks.

In this article, we discuss the theoretical foundations of these doubles in
detail, demonstrating their emergence as a surprisingly subtle consequence
of the Born-Oppenheimer approximation, and emphasizing their bundle-theoretic
and topological underpinnings. Starting with certain ``low energy''
effective theories in which classical symmetries like parity and time 
reversal are anomalously broken on quantization, we show how these
symmetries can be restored by judicious inclusion of ``high-energy'' 
degrees of freedom. This mechanism of restoring the symmetry naturally
leads to the aforementioned doublet structure. A novel by-product of this
mechanism is the emergence of an approximate symmetry (corresponding to the
approximate degeneracy of the doubles) at low energies which is not evident
in the full Hamiltonian. We also discuss the implications of this mechanism
for Skyrmion physics, monopoles, anomalies and quantum gravity.

\end{abstract}

\section{Introduction}
\subsection{The Born-Oppenheimer Approximation}
The degrees of freedom in many systems of physical interest naturally 
separate into distinct groups organized by their time scales.When this 
happens,efficient methods of calculation can also be frequently devised based 
on the idea that the time dependence of the slow variables can be ignored in
the leading approximation while treating the dynamics of the fast variables.

Molecular physics has many examples with such a sharp differentiation among 
degrees of freedom \cite{Herzberg2,ShaWil,mead}. Thus the nuclear motion in
molecules is a good deal slower than electronic motion, and the former can
be treated as slow and the latter as fast. The Born-Oppenheimer (B-O)
approximation \cite{ShaWil,mead} takes advantage of this circumstance by
imagining the nuclei to be static when solving for the electronic energy
levels. An effective nuclear Hamiltonian ${\hat{\cal H}}_S$ ($S$ for slow)
is then got from the expectation value of the exact Hamiltonian ${\cal
H}_S$ in the electronic state, the leading approximation to energy being
the eigenvalue of ${\hat{\cal H}}_S$.  

Similar examples can be found in collective nuclear models as well
\cite{BohMot2,RinSch,lesh,lech}. Thus there exist nuclei with slow-moving
cores and single-particle excitations over these cores, time scales for the
latter being shorter than those for the former. So here too, approximations
like that of Born and Oppenheimer can be a satisfactory leading
representation of reality. 

A third context, of particular interest to particle physicists, concerns 
bound states of heavy and light quarks \cite{bava}. The ratio of $b$ to $u$
quark mass  
is without doubt large while even the $c$ to $d$ mass ratio is not negligible 
by the standards of high energy theory. For example, the value of $N _c$ in 
the $1/N _c$ expansion is 3 whereas the above ratio for constituent masses 
is approximately 3 to 5. Now we can be confident that quarkonia with several 
heavy and light quarks will eventually be found, at least when these quarks 
have long enough lifetimes. It is reasonable to expect that these bound states 
will share features with molecules and nuclei mentioned above because of 
the time scales separating heavy and light quarks. That being so, it is 
plausible to imagine that a Born-Oppenheimer approximation or a variant 
thereof would be a useful first description of these systems.

\subsection{Quantum Theory of Shapes}
The low-energy bands in molecular physics and collective nuclear models are 
associated with the rotation of the nucleus as a rigid body. That is to say, 
we imagine that the configuration space $Q$ of the nucleus is got by applying 
rotations to a standard shape $S_0$ with its center-of-mass at the origin,
$S_0$ being a rigid body with a fixed fiducial orientation relative to a 
space-fixed reference frame. For methane $(C_2H_4)$ with its standard bonds 
for example, $S_0$ can be defined as the nucleus with its center-of-mass at 
the origin, carbon atoms on the third axis and the hydrogen atoms in the 2-3 
plane. $Q$ is the orbit of $S_0$ under $SO(3)$ and so is the coset space 
$SO(3)/H$, $H$ being the stability group of $S_0$. The molecular physicist 
calls $H$ the ``symmetry group'' of $S_0$. It follows that the quantum 
mechanics of nuclei treated as rigid bodies is the quantum mechanics on 
configuration spaces  $Q = SO(3)/H = SU (2)/H^*$, $H^*$ being the double 
cover of $H$. By systematically specifying all subgroups $H^*$ of $SU(2)$, 
and all qualitatively distinct quantum theories on $SU(2)/H^*$, we can also 
understand the nature of possible nuclear energy eigenstates and examine their
properties.

It is convenient henceforth to think of the double cover $H^*$ as the 
`symmetry or stability group' of the rigid body, such as the nucleus being 
considered. It was denoted by $H$ in \cite{bava}.

In previous work \cite{basiwi}, we studied quantizations of shapes
which we shall understand to be rigid bodies with configuration spaces
$Q=SU(2)/H^*$. They are topologically non-trivial and admit several of the
striking features we nowadays frequently encounter in quantum field
theory. For example, even though functions on $Q$ are tensorial fields and 
do not flip sign under $2 \pi$ rotation, there is still the possibility of
quantization with spinorial wave functions, a feature reminiscent of Skyrmion
physics \cite{BMSS}. For many molecules, $H^*$ is discrete, and in these
cases, ${\pi}_1(Q)=H^*$. In \cite{basiwi}, we concentrated on just such
molecules and their quantum physics. It is then well-known
\cite{BMSS,giulini} that there is a distinct quantum theory for each
unitary irreducible representation (UIR) of ${\pi}_1(Q)$. Wave functions in
the domain of the Hamiltonian \cite{ReeSim} in one such theory are
obtained from smooth sections of the vector bundle associated with its
UIR. These theories very nicely show the rich topological and physical  
effects of twisted bundles with flat connections in perfectly realistic 
and long familiar systems. 

\subsection{On Quantum Shapes Violating $\cal P$ and $\cal T$}
In \cite{basiwi}, we focussed attention on the effects of ${\pi}_1(Q)$ on 
parity $\cal P$ and time reversal $\cal T$. It was found that quantum theories 
of shapes can violate $\cal P$ and $\cal T$. The mechanism is much the same as 
the one leading to $\cal P$ and $\cal T$ violation in QCD in the presence of 
the theta term(for $\theta \neq 0,\pi$): $\cal P$ and $\cal T$ change the UIR 
of $H^*$ to its complex conjugate.In QCD, the analogous result is that $\cal P$
and $\cal T$ change the UIR $n \rightarrow e^{i n \theta}$ of ${\bf Z}$ to its 
complex conjugate, ${\bf Z}$ being the fundamental group of the gluon field 
configuration space. 

It merits emphasis that $\cal P$ and $\cal T$ violation being discussed here is
quantum mechanical. The left-right distinction found here is not the same
as the distinction between isomeric nuclei. It cannot be seen by a classical 
physicist. In a similar way, the QCD $\theta$ has no classical consequence and 
affects only quantum theory.

In molecular physics, there is no known microscopic source of $\cal P$ or
$\cal T$ violation. For this reason, in ref \cite{basiwi}, it was
speculated that in a more exact treatment, there must exist mechanisms
mixing states mapped to each other by $\cal P$ and $\cal T$.

\subsection{And on How Broken Symmetries are Mended: ${\cal P,T}$
Doubles and Emergent Chiral Symmetry}
In this paper, we establish that $\cal P$ and $\cal T$ are restored in a 
better treatment in a remarkable and interesting manner. Actually $\cal P$ 
and $\cal T$ violation can also happen for rigid bodies with non-discrete 
$H^*$. There are distinct quantum theories associated with distinct UIR's of 
$H^*$ here too. As our treatment works equally well for any sort of $H^*$, we 
will present the arguments without requiring its discreteness. It must be 
mentioned that the work on non-discrete $H^*$ with emphasis on topology 
was initiated before by Anezeris, Gupta and Stern \cite{ags} and that we
will gratefully use their results.

The mathematical account of the mechanism restoring $\cal P$ and $\cal T$
in molecular and nuclear physics is as follows. Let us assume that the 
domain $V^{(\rho_0)}$ of the total Hamiltonian ${\cal H}={\cal H}_S+{\cal
H}_F$ is associated with the trivial representation ${\rho}_0$ and harms
neither $\cal P$ nor $\cal T$. ${\cal H}_F$ here is the Hamiltonian of the
fast variables $F$ (or the 'fast' Hamiltonian) and ${\cal H}_S$ the
Hamiltonian of the slow variables $S$ (or the 'slow' Hamiltonian). An
eigenstate ${\psi}^{({\overline{\rho}})}_F$ of ${\cal H}_F$ is a section of
a vector bundle over $Q$ in the B-O approximation (the superscripts on the
wave functions indicate the UIR) and it can happen that this bundle is
twisted and is associated with a UIR $\overline{\rho}$. Standard
results on Berry phase \cite{berry} demonstrate this possibility. The
B-O slow Hamiltonian is not ${\cal H}_S$, it must be obtained by  averaging
$\cal H$ over ${\psi}^{({\overline{\rho}})}_F$, and when that is done, the
emergent slow Hamiltonian ${\hat{\cal H}}_S$ contains a connection  and has
a domain associated with the UIR $\rho$, the complex conjugate of
$\overline{\rho}$. So an eigenstate ${\psi}^{({\rho})}_S$ of ${\hat{\cal
H}}_S$ corresponds to $\rho$ and the product wave function
${\psi}={\psi}^{({\rho})}_S\; {\psi}^{({\overline{\rho}})}_F$ corresponds to
$\rho \otimes \overline{\rho}$. But $\cal H$ and ${\cal H}_S$ act on the
total wave function and their domain can only correspond to ${\rho}_0$.That
is now easily arranged as ${\rho}_0$ occurs in the reduction of $\rho
\otimes \overline{\rho}$. The correct total wave function in the B-O
approximation is thus the projection ${\chi}^{({\rho}_0)} = {\bf
P}[{\psi}^{({\rho})}_S \;{\psi}^{({\overline {\rho}})}_F]$ of $\psi$
to $V^{({\rho}_0)}$.  If the parity transform ${\rho}_{\cal P}$ of $\rho$
is $\overline {\rho}$ and hence that of $\overline {\rho}$ is $\rho$, the
parity transform ${\cal P} {\chi}^{({\rho}_0)}$ of ${\chi}^{({\rho}_0)}$ is
of the form ${\bf P}[{\psi}^{({\overline{\rho}})}_S \;{\psi}^{({\rho})}_F]
\in V^{({\rho}_0)}$. It is still in the domain of $\cal H$ and ${\cal
H}_S$, so there is no question of $\cal P$-violation. The same goes for
$\cal T$. But there is a doubling of states. The doubles with definite
$\cal P$, for example, in the leading approximation are linear combinations of
${\chi}^{({\rho}_0)}$ and ${\cal P}{\chi}^{({\rho}_0)}$. 

In this manner, we can see that when shapes violate $\cal P$ or $\cal T$, then 
we may have $\cal P$ or $\cal T$ doubles and no $\cal P$ or $\cal T$ symmetry 
breakdown after fast variables are included.

There is a simple and vivid manner to understand the physical mechanism 
behind these doubles. Thus consider for example a molecule like 
$N_2O$ \cite{Herzberg2}.It is a linear molecule with $O$ at one end and can be
approximated by a unit vector $\vec{n}$ (parallel to the molecule and with
the tail at $O$) when finding the rotational levels. The electronic
Hamiltonian ${\cal H}_F$ in the B-O approximation is diagonalized by
treating $\vec{n}$ as fixed. Now the system as a whole is rotationally
invariant, so for fixed $\vec{n}$, ${\cal H}_F$ is invariant under
rotations about the axis $\vec{n}$. If $\vec{L}_F$ is the fast variable
angular momentum, an eigenstate of ${\cal H}_F$ can be associated with a
definite value of $\vec{n}.\vec{L}_F$. It need not be zero, indeed it
will not be so for an odd number of electrons, as then no component of
$\vec{L}_F$ has zero eigenvalue. But $\vec{n}.\vec{L}_F$ reverses under
parity $\cal P$, and $\cal P$ is a symmetry of ${\cal H}_F$, so there is
another state with the same energy and opposite value of 
$\vec{n}. \vec{L}_F$ when the latter is non-vanishing. When we pass beyond
the B-O approximation, the exact Hamiltonian $\cal H$ mixes these levels,
thus creating mutually split even and odd energy eigenstates. 

Now of course there are many shapes in nature, and not just those described by 
$\vec{n}$. What is necessary for $\cal P$ doubles is that the shape is
reflection-asymmetric and singles out an axis. For example, if the molecule
is a pyramid with symmetry ${\bf Z}_{2N} \subset SU(2)$ around an axis
$\vec{n}$ \cite{Hammermesh}, then an eigenstate of ${\cal H}_F$ can be
associated with a definite value of $\exp[{(2 \pi
i\vec{n}.\vec{L}_F)/N}]$. [It defines $\overline{\rho}$]. Helicity is 
defined only mod $N$, $\vec{n}$ being an $N$-fold axis. Nevertheless, since
parity inverts the above exponential [and so maps $\overline{\rho}$ to
$\rho$], there are parity doubles unless $\exp[{(2 \pi i
\vec{n}. \vec{L}_F)/N}] = \pm 1$, that is, unless $\overline{\rho}=\rho$. 

Parity doubles are also $\cal T$- doubles.That is because $\cal T$ reverses 
$\vec{L}_F$ and hence $\vec{n}.\vec{L}_F$, just as $\cal P$ does.

Staggered confirmations in molecular physics are reflection-invariant. They 
cannot yield $\cal P$- doubles. They can nevertheless give $\cal T$- doubles 
both with the same $\cal P$ value, as we have discussed elsewhere
\cite{basiwi}. Such doubles will occur if $\overline{\rho} \neq \rho$. 

\subsection{About What is New, What is Predicted, and What
Experiments Confirm} 
The $\cal P$- and $\cal T$- doubles discussed above are ${\it exactly}$
degenerate in the leading B-O approximation. This exact degeneracy can only
be apparent, as there must be transitions between them in a better
calculation splitting their energies and leading, for $\cal P$-doubles, to
$\cal P$-even and $\cal P$-odd energy eigenstates. This splitting must be
small so long as the B-O approximation is decent. We are thus led to
suspect the presence of approximately degenerate $\cal K$-doubles, where
$\cal K$ is either $\cal P$ or $\cal T$.

Such parity [and perhaps also time reversal] doubles exist in 
molecular physics. The ammonia maser is based on such a double. 
They occur as well in nuclear physics for odd $A$ in the range 219-229
\cite{lesh,lech}. Elsewhere \cite{bava}, we have
suggested that doubles of this sort can occur among baryons, Skyrmions and
also composites containing grand unified monopoles. We will recall our
suggestions in particle physics later on in this work. But our primary
concern in this paper is with the bundle-theoretic aspects of these
doubles. While the computational basis of these doubles is
well-understood by chemists and nuclear physicists, their bundle-theoretic
significance is not even mentioned in their traditional literature. It is
only in modern times, after growth of interest in Berry's phase, that
Moody, Shapere and Wilczek \cite{ShaWil} elucidated formal aspects of the
B-O approximation. Our work further elaborates on these aspects, with
special emphasis on $\cal P$- and $\cal T$- symmetry restoration by the
fast variables and the attendant emergence of $\cal P$- and $\cal T$-
doubles, and covers also twisted bundles of any rank over the slow
variable configuration space. We do not think that $\cal P$ and $\cal
T$ restoration and the doubles emergent therefrom, or the higher rank
bundles have been discussed with adequate gravity before.  

Summing up, we intend to review known results in current bundle- and 
domain-theoretic language, extend them in novel directions, emphasize their 
striking features and connect them to known phenomena in quantum field theory 
like the QCD $\theta$. The new language has power and generality, and adapted 
for research in original directions. Indeed, using the experience gained from 
the foregoing considerations with $SU(2)/H^*$ as the ``slow'' manifold, we 
will be able to develop the general bundle-theoretic setting for the B-O 
approximation when $Q$ is any manifold and not limited to $SU(2)/H^*$. This 
work will then also be helpful in studying rather new applications of these 
ideas in Skyrmion physics, quantum gravity and elsewhere. All these matters 
will be taken up for consideration later on in this paper.

\subsection{A Speculation about Macroscopic Chirality}
In \cite{basiwi}, we had speculated on a scenario for the emergence of 
macroscopic classical left-right asymmetries in nature from the physics of 
these doubles. At the level of individual doubles, their classical 
description gives no clue about their difference. Nevertheless, it was 
suggested that because of quantum effects, their aggregates can lead to 
macroscopic chirality as follows: it can happen that chiral molecular samples 
having a dominant admixture of molecules with wave functions 
$\psi^{(\rho)}_S \psi^{(\overline{\rho})}_F$  or 
$[\psi^{(\overline{\rho})}_S \psi^{(\rho)}_F] \in V^{(\rho_0)}$ are 
formed in nature. They can get formed, although they are not energy 
eigenstates, for example by the action of polarized morning sunlight. If their 
decay time into energy eigenstates is sufficiently large, they can unite and 
bind into larger chiral molecules among themselves or with other neighbouring 
atoms or molecules. These bigger chiral structures would be expected to live 
longer too because of their size and may very well display macroscopic 
classical chirality. The latter is stimulated in this proposal by the
microscopic quantum mechanical $\cal P$- asymmetry. They may even be
stable. In this way, we can conceive of the formation of macroscopic chiral
structures like those actually encountered in nature.

The above scenario is speculative, but conceivable. Thus in the ammonia 
molecule for example, the parity doubles are separated by 1 cm$^{-1}$
while the rotational energies are of the order 20 cm$^{-1}$. Their ratio of
$\sim 0.05$ would also be approximately the ratio of the nuclear rotation
period to the lifetime of the chiral state, so the latter ratio is large. There
are numerous molecular parity doubles with similar large numbers. It is now
an interesting matter, open for investigation, to confirm or deny this
scenario by detailed calculations.  

\subsection{Contents of the Paper}
Quantization of shapes in modern language, and the demonstration that they can 
violate $\cal P$ and $\cal T$ by quantum mechanical effects, have been
fully covered in \cite{basiwi}. We will briefly summarize the results of
that reference in Section 2. In Section 3, we take up the treatment of the
fast variables for the ``slow'' configuration space $Q=SU(2)/H^*$ and
derive the 
general form of the eigenfunctions ${\psi}^{({\overline{\rho}})}_F$ of
${\cal H}_F$ as sections of a vector bundle over $Q$. The bundle is
associated with the UIR ${\overline{\rho}}$ of $H^*$. The fast variables
$F$ can describe electrons in molecular physics and ``intrinsic
components'' \cite{BohMot2} or excitations over the core in nuclear
physics. Such details are not important at our level of generality. In
Section 4, we derive the effective Hamiltonian ${\hat{\cal H}}_S$ for the
above $Q$. It now contains a connection induced by the fast variables. The
connection determines the bundle appropriate for the slow variables. This
bundle is associated with the representation $\rho$ complex conjugate to
${\overline{\rho}}$. The nature of the total wave function claimed earlier
is in this manner established. There is no $\cal P$ or $\cal T$ violation
in the total system even though there could be such violation in the
quantum theory of just the shapes if $\rho \neq {\overline{\rho}}$.

Calculations of the sort we perform here for the derivation of 
${\hat{\cal H}}_S$ have appeared elsewhere before \cite{moshwi,ShaWil}.

In Section 5, after recapitulating the nature of the total wave functions and 
how $\cal P$ and $\cal T$ are restored as symmetries, we explain a standard 
mechanism mixing the wave functions ${\psi}^{({\rho})}_S {\psi}^
{({\overline{\rho}})}_F$ and ${\psi}^{({\overline{\rho}})}_S 
{\psi}^{({\rho})}_F$. It is this which splits the parity and time reversal 
doubles in the leading order. 

Section 6 discusses the possibility of $\cal P$- and $\cal T$- doubles in 
quark physics and argues in particular that they may occur in $cbu$ or $cbd$ 
systems. The difference between the conventional quark model states and these 
B-O states is also pointed out. Certain other areas of interest to particle 
physics where the Born-Oppenheimer ideas may be fruitful are postponed for 
discussion to Section 8.

Section 7 develops the theory of the B-O approximation when $Q$ is not just 
$SU(2)/H^*$, but a more general manifold. The concluding Section 8 indicates 
how Section 7 can be applied to Skyrmion and monopole physics, anomalies and 
quantum gravity. Work on all these topics is in progress with promising 
results. 

\section{When do Shapes Violate $\cal P$ and $\cal T$?}

The total Hamiltonian $\cal H$ of the slow and fast variables can be 
written as the sum of two terms:

\begin{equation}
	{\cal H} = {\cal H}_S + {\cal H}_F.
\end{equation}
In this decomposition, ${\cal H}_S$ depends only on the slow degrees of 
freedom. Its typical form when $S$ is a shape with configuration space 
$SU(2)/H^*$ is 

\begin{equation}
	{\cal H}_S = \vec{L}_S^2/2{\cal I},
\label{H_S}
\end{equation}
$\vec{L}_S$ being the angular momentum of $S$ and ${\cal I}$ its moment of
inertia. The Hamiltonian ${\cal H}_F$ contains the fast degrees of freedom
and also the interaction between $S$ and $F$.  

{\it The discussion in this section concerns ${\cal H}_S$}. It is important to 
recognize that ${\cal H}_S$ is not the B-O slow Hamiltonian ${\hat{\cal H}}_S$
emergent from the average of $\cal H$ over ${\psi}^{({\overline{\rho}})}_F$. 
The distinction is of particular significance for nonabelian 
${\overline{\rho}}$. We will resume further examination of ${\hat{\cal H}}_S$ 
and ${\cal H}_S$ in later sections. 

The symmetry groups $H^*$ of shapes can be discrete or continuous. Our 
discussion applies to all choices of $H^*$. For a classical shape with 
configuration space $Q=SU(2)/H^*$, there is a quantum shape for each UIR 
$\rho$ of $H^*$ \cite{basiwi,BMSS,giulini}. Wave functions in the domain of the
Hamiltonian are sections of the vector bundle over $Q$ associated with the
representation $\rho$. They can be found as follows \cite{sorkin}. 

Let [$\rho$] be the dimension of the UIR $\rho$ and $\rho (h)$ the matrix 
of $h \in H^*$ in $\rho$. The above wave functions are constructed from 
smooth vector-valued functions $f$ on $SU(2)$ with ``internal'' dimension 
[$\rho$],  

\begin{equation}
 	f=(f_1, f_2,....f_{[\rho]}), \quad f_j(g) \in {\bf C}\;{\rm for}\;g
 		\in SU(2), 
\label{wf}
\end{equation}
which also have the transformation property 

\begin{equation}
 	f_j(gh) = f_k(g){\rho}_{kj}(h), \quad h \in H^*.
\end{equation}

\subsection{Discrete $H^*$}
When $H^*$ is discrete, the fibres of the principal bundle $H^* \rightarrow 
SU(2) \rightarrow SU(2)/H^*$ are also discrete. In this case, the Hamiltonian 
${\cal H}_S$ can be defined as a self-adjoint operator on the wave
functions (\ref{wf}) for any choice of $\rho$. This is because the connection
in the above bundle is flat \cite{basiwi,BMSS,giulini,sorkin}. There is
thus a certain freedom in the choice of the domain for ${\cal H}_S$ with
its attendant quantization ambiguity mentioned above.   

Discrete symmetry groups occur for pyramids, tetrahedra, and eclipsed 
conformations (with dihedral symmetry groups). For these shapes, it was
shown in \cite{basiwi} that $\cal P$ or $\cal T$ violation will occur if and
only if $\rho \neq {\overline{\rho}}$. In contrast, staggered confirmations
(also with dihedral symmetry groups) never violate $\cal P$ while they
continue to violate $\cal T$ if $\rho \neq {\overline{\rho}}$. Hence they
can also violate $\cal PT$. 

The cube and the dodecahedron, and their duals, the octahedron and 
icosahedron, also have discrete symmetry groups, but they cannot violate 
$\cal P$ or $\cal T$ \cite{basiwi}. 

\subsection{The Case $H^* = U(1)$}
$SU(2)$ has a $U(1)$ subgroup which is unique up to conjugation. It can be 
taken to be 
\begin{equation}
	U(1) = \{ e^{i {\tau}_3 \theta/2} : 0 \leq \theta \leq 4 \pi \} , 
		\quad {\tau}_i = {\rm Pauli \; matrices}. 
\end{equation}
The space $Q=SU(2)/U(1)$ is the two-sphere $S^2$ of unit vectors or arrows 
with heads \cite{BMSS}:

\begin{equation}
	S^2 = \{ \vec{n}:\vec{n} \in {\bf R}^3, 
		\vec{n}.\vec{n}=1 \}.
\end{equation}
The stability group of $\vec{n}$ is $U(1)$. The latter is the 
stability group of any shape with axial symmetry. Any such shape has the above
 configuration space $Q$. 

The UIR's $\rho = {\rho}_K (K \in {\bf Z}/2)$ of $H^* = U(1)$ are given by 

\begin{equation}
	{\rho}_K:e^{i {\tau}_3 \theta/2} \rightarrow {\rho}_K (e^{i
		{\tau}_3 \theta/2}) = e^{i K \theta}.
\end{equation}
The functions (\ref{wf}) for $\rho = {\rho}_K$ give sections of $U(1)$
bundles over $S^2$ for Chern class $K$. These are the bundles for
charge-monopole systems with $K = eg/{4 \pi}$, $e$ and $g$ being the 
electric and magnetic charges \cite{BMSS}. Now just as for the latter,
$\cal P$ and $\cal T$ change ${\rho}_K$ to ${\rho}^*_K$, or $K$ to $-K$,
and hence are violated for $K \neq 0$. But $\cal PT$ is always good. 

When $H^* = U(1)$, the Hamiltonian ${\cal H}_S$ itself completely determines 
the bundle, that is, the choice of $K$. This is because the bundles for 
$K \neq 0$ do not admit flat connections. The Hamiltonian ${\cal H}_S$ is not 
(\ref{H_S}), but rather has the general form

\begin{equation}
 {\vec{\cal L}_S}^2 /2{\cal I},
 \label{H_SBO}
\end{equation}
$\vec{\cal L}_S = \vec{L}_S$ + a term involving the connection appropriate
for ${\rho}_K$. For example, $\vec{\cal L}_S$ can be $\vec{r} \times 
(\vec{p}-e \vec{A})$ where $\vec{r}$ and $\vec{p}$ are the relative
coordinate and its conjugate momentum, $\vec{A}$ is the vector potential of
the monopole field $(g/{4 \pi}) \vec{r}/r^3$ and the entire expression is
restricted to the two-sphere $\vec{r} = \vec{n}$. 

\subsection{The case $H^* = D^*_{\infty}$}
$SU(2)$ has yet another non-discrete subgroup which is also unique up to 
conjugation. It is the `infinite' dihedral group 
\begin{equation}
  	D^*_{\infty} = \{ e^{i \tau_3 \theta/2}, i \tau_2:\quad 0\leq
  		\theta \leq 4 \pi \}, 
\label{dihedral}
\end{equation}
$i {\tau}_2$ being rotation by $\pi$ around the second axis. The shape 
$Q= SU(2)/D^*_{\infty}$ is the projective two-sphere of headless arrows:
\begin{equation} 
	Q={\bf RP}^2,
\end{equation}
the headless arrow having $D^*_{\infty}$ (or rather, a group isomorphic to
(\ref{dihedral})) as its stability group. 
The hydrogen molecule is an example described by a headless arrow and the 
configuration space ${\bf RP}^2$.

The one-dimensional UIR's of $D^*_{\infty}$ are the trivial one (which need 
not be discussed further) and the representation
\begin{eqnarray}
	\rho: e^{i \tau_3 \theta/2} &\rightarrow & \; 1, \nonumber \\
	      i \tau_2 &\rightarrow &-1. \label{1dUIR}
\end{eqnarray}
It also has the two-dimensional UIR's

\begin{eqnarray}
      {\rho}_K: e^{i \tau_3 \theta/2} &\rightarrow {\rho}_K(e^{i \tau_3 
		\theta/2}) &= \left[ \begin{array}{cc}
					e^{i K \theta} & 0 \\
					0 & e^{-i K \theta} \\	
				   \end{array}
		    	     \right],  \\
	  i \tau_2 &\rightarrow {\rho}_K(i \tau_2) &= i \tau_2, \\
	  K & \in & {\bf N}/2.
\end{eqnarray} 

The group ${\rho}(D^*_{\infty})$ is the discrete group ${\bf Z}_2$. In this
case, just as for the discrete $H^*$, there is a two-fold ambiguity in
quantizing ${\cal H}_S$: we can use either the trivial UIR or the UIR 
(\ref{1dUIR}).

Let $|m  \rangle, \; m  \in \{ j,-j+1,...j\}$ be the standard orthonormal
basis for the (2$j$+1)-dimensional UIR of $SU(2)$ with the 3rd component
$J^{(j)}_3$ of angular momentum being diagonal:

\begin{eqnarray}
	J^{(j)}_3|m \rangle & = m|m \rangle, \\
	\langle m'|m \rangle & = {\delta}_{m'm}. 
\end{eqnarray}

Choose $j$ so that $\pm K$ occur in the spectrum of  $J^{(j)}_3$. Then the 
restriction of 
$$
	e^{i J^{(j)}_3 \theta}, \quad e^{i J^{(j)}_2 \pi}
$$
to the subspace spanned by  $|K \rangle$ and $|-K \rangle$ gives the UIR 
${\rho}_K$, $J^{(j)}_2$ here being the second component of angular momentum. 

A headless arrow is invariant under reflection. So we can set ${\cal P}= 1$ 
on wave functions and there is no parity violation for any of the UIR's of 
${\bf RP}^2$. 

There is no $\cal T$ violation either for any of the UIR's of
${\bf RP^2}$. That is because the one-dimensional UIR's are real while the
UIR's $\overline{\rho}_K$ complex conjugate to ${\rho}_K$ is equivalent to
${\rho}_K$: 

\begin{equation}
	\overline{\rho}_K(h) = {\rho}_K (i {\tau}_2) {\rho}_K (h)
		{\rho}_K^{-1} (i {\tau}_2), h \in D^*_{\infty}.
\end{equation}

The UIR $\rho_K(D^*_{\infty})$ is not discrete. So the option of using this 
representation is available only if the Hamiltonian itself is of the form 
(\ref{H_SBO}). Now as ${\rho}_K$ is two-dimensional, it is a $2 \times 2$
matrix of differential operators. We will discuss such operators further in
Section 4. 

\section{The Fast Wave Function}

As mentioned in the Introduction, until Section 7, we will specialize to the 
case $SU(2)/H^*$ for the slow configuration space $Q$. We will also 
occasionally explain or illustrate a point using molecules as they provide 
excellent examples for these $Q$.

A fast wave function for us means an energy eigenstate for the Hamiltonian 
${\cal H}_F$ in the B-O approximation. The slow variables of the shape are 
looked upon as static in this approximation.

Let us assume that the total Hamiltonian $\cal H$ for the interacting fast 
and slow variables is rotationally invariant. This assumption can be avoided 
as shown in Section 7, but is realistic, so let us keep it for now.

Let $S_0$ be the conveniently chosen standard shape. The group $SU(2)$, the 
two-fold cover of $SO(3)$ of spatial rotations, acts transitively on $Q$. 
Denoting this action by $S_0 \rightarrow gS_0$, $g \in SU(2)$, any shape $S$ 
can be written as $gS_0$.

Let $H^*$ be the subgroup of $SU(2)$ leaving $S_0$ invariant. Then the 
invariance group of $gS_0$ is $gH^*g^{-1}$. It is isomorphic to $H^*$, but 
not identical to $H^*$ when regarded as a subgroup of $SU(2)$. 

The group $SU(2)$ acts on the operators ${\cal O}_F$ intrinsic to the fast
system and on the eigenstates $|.\rangle$ of ${\cal H}_F$.Let us denote
these actions by ${\cal O}_F \rightarrow U(s) {\cal O}_F U(s)^{-1}$ and
$|.\rangle \rightarrow U(s)|. \rangle,$  $s \in SU(2), U(s)$ being a unitary
operator. Let $\vec{L}_F$ be the generators of this $SU(2)$. They are the
fast angular momenta.   

As $\cal H$ is invariant under $SU(2)$ and $hS_0 = S_0 \;{\rm for}\; h \in
H^*$, ${\cal H}_F$ is invariant under the $H^*$ action on $F$ alone when
the slow variable has the configuration $S_0$. Eigenfunctions of ${\cal
H}_F$ for a fixed eigenvalue will therefore transform by a representation
of $H^*$. This representation will be UIR unless there is ``accidental''
degeneracy as for the non-relativistic hydrogen atom. Results for a general
representation, which will be a direct sum of UIR's, in any case follow from
those for the latter, so let us assume the representation to be a UIR
$\overline{\rho}$.
 
Suppose then that the shape is $S_0$ and that for that shape,
\begin{equation}
\{ |m \rangle = m \in {\rm a\; suitable\; index\; set\;} \overline{I} \}
\end{equation}
is an orthonormal basis spanning an eigenstate of ${\cal H}_F$ and carrying 
also the UIR $\overline{\rho}$: 

\begin{eqnarray}
	{\cal H}_F|m \rangle & = & \epsilon \;|m \rangle, \nonumber \\
	U(h)|m \rangle & = &  |m' \rangle \overline{\rho}(h)_{m'm}, \\
	\label{estateH_F}
	\langle m'|m \rangle & = & {\delta}_{m'm}. \nonumber
\end{eqnarray}

The corresponding states when the shape is $gS_0$ are just $U(g)|m \rangle$:

\begin{eqnarray}
	{\cal H}_F U(g)|m \rangle &=& \epsilon \; U(g)|m \rangle, \nonumber \\
	\langle m'|U(g)^{\dagger} U(g)|m \rangle &=& {\delta}_{m'm}.
\end{eqnarray}
As shown here, the new states are orthogonal since $U(g)$ is unitary. [The
background shape in ${\cal H}_F$ here is $gS_0$ while it is $S_0$ in
(\ref{estateH_F}). Still, for notational simplicity, we have denoted this
operator by the same symbol ${\cal H}_F$ used in (\ref{estateH_F}) although
this is not quite correct.]

The eigenvalue $\epsilon$ has no shape dependence because of the assumed 
$SU(2)$ invariance of ${\cal H}_F$.

A fast wave function depends not only on $g$ and $m$, but also on variables
$\{ {\xi}_{\beta} \}$ describing the fast configuration in the body-fixed 
frame of the slow variables. They are invariant under $SU(2)$, being the 
analogues of radial coordinates for a particle in a central potential. It is 
convenient to rewrite $|m \rangle$ in a particular way where this and angular 
momentum dependence can be made explicit. 

So let us expand $|m \rangle$ in a basis $|m,j \rangle$ where 
$\vec{L}_F^2$ is diagonal:

\begin{eqnarray}
	\vec{L}_F^2 |m,j \rangle&=& j(j+1)|m,j \rangle, \nonumber \\
	\langle m',j'|m,j \rangle &=& {\delta}_{m'm} {\delta}_{j'j}.
\end{eqnarray}
The expansion is 

\begin{eqnarray}
	|m \rangle &=& {\sum}_j |m,j \rangle {\alpha}_j, \nonumber\\
	{\alpha}_j  : \{ {\xi}_{{\beta}} \}& \rightarrow &{\alpha}_j(\{
	{\xi}_{{\beta}} \}) \in {\bf C}.
\label{redvar}
\end{eqnarray}
Of course a $j$ would occur in the sum here only if the spin $j$ UIR of
$SU(2)$ on restriction to $H^*$ contains $\overline{\rho}$.

Let $d \mu (\{ {\xi}_{\beta} \})$ be the measure of integration for the
scalar product of functions of $\{ {\xi}_{\beta} \}$. Then the
orthonormality of $|m \rangle$ gives 

\begin{equation}
	\int d \mu (\{ {\xi}_{\beta} \}) \; {\sum}_j {\mid {\alpha}_j(\{ {\xi}_
		{\beta} \}) \mid}^2 = 1.
\end{equation}

We can expand ${\alpha}_j$ in a complete set of functions orthonormal for the 
measure $d \mu (\{ {\xi}_{\beta} \})$ if desired. 

A basis $\{ |k,j \rangle \}$ of $(2j+1)$ orthonormal states spanning
the spin $j$  
UIR of $SU(2)$ can be chosen so that $\{ |m,j \rangle: m \in \overline{I}
\} \subset \{ |k,j \rangle \}:$

\begin{eqnarray}
	\langle k',j'|k,j \rangle & = {\delta}_{k'k} {\delta}_{j'j},\nonumber\\
	\{ |m,j \rangle:m \in \overline{I} \} & \subset \{ |k,j\rangle \}.
\end{eqnarray}
In this basis, 

\begin{equation}
	U(s)|m,j \rangle =|k,j \rangle D^j_{km}(s), m \in \overline{I}
	\label{3.8}
\end{equation}
where $k$ is summed over $(2j+1)$ values and $D^j(s)$ are the rotation 
matrices in the chosen basis. Thus we get the useful formula

\begin{equation}
	{\chi}^F_m(g,.) \equiv U(g)|m \rangle = \sum |k,j \rangle D^j_{km}(g)
		{\alpha}_j(.).
\label{fastwf} 
\end{equation}

A property of significance is the following transformation of ${\chi}^F_m$ 
implied by (\ref{fastwf}):

\begin{equation}
	{\chi}^F_m(gh,.) = {\chi}^F_{m'}(g,.) \overline{\rho}(h)_{m'm}.
\label{fastwftrn}
\end{equation}

\section{The Slow Wave Function}
The effective ``slow'' Hamiltonian ${\hat{\cal H}}_S$ for the slow degrees of 
freedom in the B-O approximation differs in general from the ``true'' slow 
Hamiltonian ${\cal H}_S$. An energy eigenstate of ${\hat{\cal H}}_S$, or 
perhaps of ${\cal H}_S$, is what we informally and indiscriminately call the 
slow wave function. It thus refers to a wave function with dependence only on 
slow variables. 

The slow variables belong to massive bodies. Their spin effects, which depend 
on powers of inverse mass, are therefore small and are ignored in the 
B-O approximation.

In molecular physics, while discussing rotational bands in the B-O 
approximation, all degrees of freedom of the nucleus except its overall shape 
are frozen. The nature of the angular momentum for the slow system is thus
of central importance in this approximation. We will make the following
assumption about it in what follows: for the Hamiltonians ${\cal H}_S$ and
${\cal H}$, and hence for the exact slow wave functions, it is just the
orbital angular momentum $\vec{L}_S$.

While discussing $\vec{L}_S$, we can put the center-of-mass of the 
slow system at the origin, and when talking about molecules distribute the 
constituent nuclei so that altogether they have the symmetry group $H^*$. We 
can for example imagine them to be connected by chemical bonds. Then 
components of $\vec{L}_S$ become just the vector fields generating 
rotations. Their domain of definition being obtained from smooth functions on 
$Q$, and with ${\cal H}_S$ being (\ref{H_S}), the bundle of slow wave
functions on $Q$ is trivial. This then is the content of our assumption
whether or not we are dealing with molecular physics.

We restrict angular momentum in this manner partly for convenience. It
could happen for a general system that the angular momentum of the (exact) 
slow system is not $\vec{L}_S$, but a twisted version thereof appropriate
for some nontrivial bundle over $Q$. But this case can be dealt with
effortlessly with minor modifications of our discussion. As for molecular
physics, when the effects of spin $\vec{S}$ of the slow constituents are
not ignored, the angular momentum is not $\vec{L}_S$, but $\vec{J} =
\vec{L}_S + \vec{S}$, $\vec{S}$ being the total spin. If the time scales
associated with the motion of spins are short, we can treat $\vec{S}$ too
as a fast variable. It could then acquire a non-zero component along the
body-fixed axis. The form of angular momentum would then be altered and
the bundle of slow wave functions over $Q$ would get twisted. In that event
we must modify the discussion somewhat, but the changes are cosmetic, as we
can include $\vec{S}$ too among fast variables. No separate discussion is
thus needed when the effects of spin of the slow system are not ignored
provided spin is fast. But that is not the case if $\vec{S}$ for example is
not fast. This situation will not be covered in this paper.

The eigenstate of $\cal H$ in the B-O approximation is assumed to have
the form

\begin{equation}
	\phi = \sum {\psi}^m_S \; {\chi}^F_m
\end{equation}
where ${\psi}^m_S$ is a function only of the slow variables. [Here
${\chi}^F_m$ depends on $g$ and $\{ {\xi}_{\beta} \}$, ${\chi}^F_m(g,.)$
being the function of $\{ {\xi}_{\beta} \}$ in(\ref{fastwf})]. 
Hence 

\begin{equation}
	{\cal H} \phi = {\cal H}_S \sum {\psi}^m_S \;{\chi}^F_m + \epsilon
		\sum {\psi}^m_S \;{\chi}^F_m.    
\end{equation}

We now average this over the fast degrees of freedom. On taking the scalar 
product with ${\chi}^F_m$, the eigenvalue problem 

\begin{equation}
	{\cal H} \phi = E \phi
\end{equation}
becomes 

\begin{equation}
	{\sum}_n ({\chi}^F_m, {\cal H}_S \;{\chi}^F_n) {\psi}^n_S +
		\epsilon {\psi}^m_S = E {\psi}^m_S
\end{equation}

In the traditional B-O approximation, only the intermediate states
${\chi}^F_j$ are retained between the two $\vec{L}_S$ when
evaluating the first term using (\ref{H_S}). In this approximation, this
kinetic energy term ${\vec{L}_S}^2 /2{\cal I}$ becomes
${\vec{{\cal L}_S}^2}/2{\cal I}$, where ${{\cal L}_S}_j$ are 
matrix-valued differential operators:

\begin{equation}
	(\vec{\cal L}_S)_{mn} = {\delta}_{mn}
 		\vec{L}_S+({\chi}^F_m,[{\vec{L}_S} \;{\chi}^F_n]). 
\label{curlyL}
\end{equation}
The square brackets here signify that $\vec{L}_S$ within it differentiates
only ${\chi}^F_n$ ( and not objects which may occur further to the right). 

There is a certain delicacy in the definition of $\vec{L}_S$ in
(\ref{curlyL}) since orbital angular momentum acts on functions on $Q$
whereas ${\psi}^m_S$ and ${\chi}^F_n$ depend on $g$. We will address this
issue later on in Section 5. 

The B-O slow Hamiltonian and the equation for energy are therefore
\cite{mead,ShaWil,moshwi}

\begin{eqnarray}
	{\hat{\cal H}}_S &=& \vec{{\cal L}_S}^2/2{\cal I} +
		\epsilon,\nonumber  \\
	{\hat{\cal H}}_S {\psi}_S &=& E {\psi}_S,
		\quad {\psi}_S =({\psi}^1_S, {\psi}^2_S,...).
\label{H_SBOfull}
\end{eqnarray}

The operator $\vec{\cal L}_S$ is not always an angular
momentum. Its components need not fulfill angular momentum commutation
relations. But we want to know the angular momentum commuting with
${\hat{\cal H}}_S$, or rather, the bundle describing the energy
levels of ${\hat{\cal H}}_S$. We can address this task as follows. 

The choice of $|m \rangle$ is not unique. We can get another choice by the 
action of an $h$ in $H^*$. Thus we can change $|m \rangle$ to $|m' \rangle
\overline{\rho}(h)_{m'm}$ and thereby also change ${\chi}^F_m$ 
according to 

\begin{equation}
	{\chi}^F_m \rightarrow {\chi}^F_{m'} \; \overline{\rho}(h)_{m'm}.
\end{equation}
This corresponds just to the change 

\begin{equation}
	{\chi}^F_m(g,.) \rightarrow {\chi}^F_m(gh,.)
\end{equation}
in the choice of basis for energy eigenfunctions of ${\hat{\cal H}}_S$.

The responses of $\vec{{\cal L}_S}$ and ${\hat{\cal H}}_S$
to the transformation (\ref{fastwftrn}) are

\begin{eqnarray}
	\vec{{\cal L}_S} &  \rightarrow \;
		\overline{\rho}(h)^{-1} \;\vec{{\cal L}_S} 
		\; \overline{\rho}(h),  \nonumber \\
	{\hat{\cal H}}_S & \rightarrow \; \overline{\rho}(h)^{-1}
		\;{\hat{\cal H}}_S \; \overline{\rho}(h).
\end{eqnarray}
The response of ${\psi}_S$ is thus

\begin{equation}
	{\psi}_S \; \rightarrow \; \overline{\rho}(h)^{-1} \;
		{\psi}_S.
\end{equation}

The group $SU(2)$ as a manifold is an $H^*$-principal bundle over
$Q=SU(2)/H^*$. It is convenient to regard ${\psi}_S$ as a function on
$SU(2)$. [The conventional wave function, which is a section of an
associated bundle can be obtained by a ``gauge choice'', that is by
restricting the above ${\psi}_S$ to a section of this principal
bundle.] It thus follows that 

\begin{equation}
	{\psi}_S(gh) = \overline{\rho}(h)^{-1} \; {\psi}_S(g).
\label{slowwftrn}
\end{equation}
As $\overline{\rho}$ is unitary, we thus get

\begin{equation}
	{\psi}^m_S(gh) = {\psi}^{m'}_S(g) \; {\rho}(h)_{m'm} 
\end{equation}
where $\rho$ is the UIR complex conjugate to $\overline{\rho}$. 
{\it So ${\psi}_S$ comes from the associated vector bundle for the UIR
complex conjugate to the UIR of the fast wave function} \cite{ShaWil}. 

The $SU(2)$ group of angular momentum acts on ${\psi}_S$ by left
multiplication on $g$. If $U(g')$ is the corresponding unitary
operator, this action explicitly is

\begin{equation}
	[U(g') {\psi}_S](g) = {\psi}_S({g'}^{-1}g).
\end{equation}
The expression for angular momentum can be written down from the
infinitesimal form of this formula. 

Note that as $g'$ acts on the left, and $h$ on the right, of $g$,
these two actions commute. 

\subsection{Emergence of Unitary Gauge Symmetries}
In the above discussion, we limited ourselves to transformations of
${\chi}^F_m$ induced by $H^*$. There is no good reason for this
restraint, we can transform them by any unitary transformation and they
will still remain orthonormal and degenerate eigenstates of ${\cal
H}_F$. We can even choose the elements of the unitary matrix to be
functions on $Q$ without spoiling these properties. Thus the general
transformation we can perform is 

\begin{equation}
	{\chi}^F_m(g,.)\; \rightarrow \; {\chi}^F_{m'}(g,.)u_{m'm}(gS_o)   
\label{gaugetrn}
\end{equation}
where we have labeled points of $Q$ by the shapes $gS_o$. These
transformations form the unitary gauge group $u_{[\overline{\rho}]}$,
$[\overline{\rho}]$ being the dimension of the UIR $\overline{\rho}$. 

Let $\vec{A}$ be the $[\overline{\rho}] \times [\overline{\rho}]$ matrix
with components  

\begin{equation}
	{\vec{A}}_{mn} = ({\chi}^F_m,[{\vec{L}_S} \; {\chi}^F_n]).
\end{equation}
Its response to the transformation (\ref{gaugetrn}) is

\begin{equation}
	\vec{A} \rightarrow u^{-1}\vec{A}u + u^{-1}[\vec{L}_S u] .
\end{equation}
This shows that $\vec{A}$ is a connection for the above gauge group. 

The transformation properties of $\vec{{\cal L}_S}$, ${\hat{\cal H}}_S$ and
${\psi}_S$ are 
\begin{eqnarray}
	\vec{{\cal L}_S} & \rightarrow
		u^{-1} \vec{{\cal L}_S} u, \\ 
	{\hat{\cal H}}_S & \rightarrow u^{-1}{\hat{\cal H}}_S u, \\
	{\psi}^m_S & \rightarrow {\psi}^{m'}_S u^*_{m'm}.
\end{eqnarray}
Thus a unitary gauge group has emerged from the B-O approximation.

The functions $u_{m'm}$ on $Q$ can be thought of as functions
${\hat{u}}_{m'm}$ on $SU(2)$:
 
\begin{equation}
	u_{m'm}(gS_0) = {\hat{u}}_{m'm}(g).
\end{equation}
They are not general functions on $SU(2)$ however, having the
invariance property

\begin{equation}
	{\hat{u}}_{m'm}(gh) = {\hat{u}}_{m'm}(g)\quad {\rm for}\quad h\in H^*.
\label{restriction}
\end{equation}
It is natural to enquire if the gauge group can be further enlarged by
dropping this restriction. The resultant gauge transformations $\{ \hat{v}
\}$ would consist of functions on $SU(2)$ with values in
$[\overline{\rho}]$- dimensional unitary matrices. They would not be 
constrained by an equation like (\ref{restriction}). So they cannot always be
regarded as functions on $Q$. 

But physics does not permit this enlargement of the gauge group. In
physics, the scalar product (.,.) between two fast wave functions, which
${\it a \; priori}$ is a function on $SU(2)$, must project down to a
function on $Q$ and thus be invariant under $g \rightarrow gh$. This is
because scalar products are observable, and it is only functions on $Q$ and
not general functions on $SU(2)$ which are observable. For it is $Q$ and
not $SU(2)$ which is the configuration space.

Scalar products between any two ${\chi}^F_m$'s do not at all depend on
$g$. We can assume that they are permissible fast wave functions. If
${\chi}^F_{m'} {\hat{v}}_{m'n}(g)$ is also a permissible wave function,
its scalar product with ${\chi}^F_m$,

\begin{equation}
	({\chi}^F_m,\;{\chi}^F_{m'} {\hat{v}}_{m'n}(g)) =
		{\hat{v}}_{mn}(g),
\end{equation}
must be a function on $Q$. So $\hat{v}$ must be a $\hat{u}$ with the
property (\ref{restriction}), and ${\cal U}_{[\overline{\rho}]}$ is the
gauge group. 

\section{Symmetry Restoration: $\cal P$,$\cal T$-Doubles and How They
Split}

\subsection{Symmetry Restoration}
We have explained before that $\cal P$ and $\cal T$ get violated by a 
quantum shape having UIR $\rho$ if $\rho$ becomes an inequivalent UIR
under their effect. For our $Q$, violation happens by $\rho$ becoming
its complex conjugate $\overline{\rho} \neq \rho$.

While quantum shapes can violate $\cal P$ and $\cal T$, we cannot entertain
the conjecture that a quantum molecule does so,its microscopic Hamiltonian
being rigorously invariant under these symmetries. So $\cal P$ and $\cal
T$, even if spoilt by shapes, must become good again after the effect of
the electronic cloud is accounted for. Let us once more explain this
remarkable symmetry restoration.  

When $\rho$ becomes $\overline{\rho}$ under $\cal P$ or $\cal T$, the
slow wave function ${\psi}_S$ becomes ${\cal K}{\psi}_S$ (where ${\cal K}$
is either $\cal P$ or $\cal T$) and transforms by $\overline{\rho}$. If
$\rho \neq \overline{\rho}$, $\cal P$ or $\cal T$ does not leave the domain
of the slow Hamiltonian invariant and spoils the symmetry. This is what we
claimed in previous work \cite{basiwi}.  

We now include the electronic cloud. In discussing this cloud, it is
necessary to clarify our assumption about the actions of  $\cal P$ or
$\cal T$ on the internal wave functions ${\alpha}_j$ which are functions
of $\{ {\xi}_{\beta} \}$. Our assumption is that this action does not cause
domain problems for the Hamiltonian. It is thus enough to pay attention to
$\rho$ and $\overline{\rho}$.  

When $\overline{\rho}$ changes to $\rho$ under $\cal K$, ${\chi}^F$ becomes
${\cal K}{\chi}^F$ and transforms by $\rho$. So while ${\psi}_S^m \;
{\chi}_m^F$ transforms by $\rho \otimes \overline{\rho}$. As for the
complete wave function $\phi = {\psi}_S^m \; {\chi}_m^F$, it becomes  
 
\begin{equation}
	{\cal K}\phi = ({\cal K}{\psi}_S)^m \; ({\cal K}{\chi}^F)_m,
\end{equation}
the first factor being associated with $\overline{\rho}$ and the second
with $\rho$. The overwhelming question is then whether $\phi$ and
${\cal K} \phi$ are in the domain of the full Hamiltonian $\cal H$. 

Our hypothesis has been that the domain $V^{({\rho}_0)}$ of $\cal H$
is associated with the untwisted bundle corresponding to the trivial
UIR ${\rho}_0$. It thus consists of appropriately smooth functions on
$Q$. Now $\phi$ is invariant under the action of $H^*$,

\begin{eqnarray}
	\phi(gh,.) &=& {\psi}^{m'}_S \rho(h)_{m'm}\;{\chi}^F_{m''}
		 	\overline{\rho}_{m''m} \\
		   &=& \phi(g,.)
\end{eqnarray}
and hence 
\begin{equation}
	\phi \in V^{({\rho}_0)}
\end{equation} 
and is a function on $Q$. This is a relief: we would not be able to proceed
further if $\phi \notin V^{({\rho}_0)}$. A similar calculation shows
that ${\cal K} \phi$ too is $H^*$-invariant. We therefore have that 

\begin{equation}
	{\cal K} \phi \in V^{({\rho}_0)}.
\end{equation}
This shows that the quantum molecule preserves ${\cal K} = {\cal P}\;
{\rm or}\; {\cal T}$ even though the quantum shape may spoil it. 

But the effective Hamiltonian $\hat{\cal H}_S$ does disturb $\cal
K$-symmetry if $\rho \neq \overline{\rho}$. There is then no sense in
applying $\hat{\cal H}_S$ to $\phi$ since the latter is not in the
domain of $\hat{\cal H}_S$ which is derived from the UIR $\rho$. 
{\it We thus see that an effective Hamiltonian can show spurious symmetry
violations, which however get restored when fast degrees of freedom are
judiciously included.}\cite{bava}. This could be a significant insight
taught to us by the B-O approximation.  

\subsection{Meaning of $\vec{L}_S$ in $\vec{{\cal L}_S}$}  
We now take up the precise definition of $\vec{L}_S$ which occurs in
(\ref{curlyL}). It is not the orbital angular momentum: orbital angular
momentum acts on functions on $Q$ whereas this $\vec{L}_S$ acts on certain
functions of $g$ which for $\rho \neq \overline{\rho}$ do not admit
interpretation as suitably smooth functions on $Q$. We should use a
different symbol for this $\vec{L}_S$, but we avoided that to prevent
excessive early perplexity in the reader, if any.

We have shown that $\phi$ is a function on $Q$. Hence $\vec{L}_S$ is
well-defined on $\phi$ and acts as orbital angular momentum. Now the action
$g \rightarrow e^{i \vec{\theta}.\vec{\tau}/2}g$ of $SU(2)$ commutes with
the action of $H^*$. So $SU(2)$ acts on 
$Q=SU(2)/H^* = \{ gH^* \}$ and there becomes its rotations. It follows that if
${\vec{L}}^{(\rho)}$ and ${\vec{L}}^{(\overline{\rho})}$ are the generators
of the above $SU(2)$ acting on the argument $g$ of ${\psi}_S$  and ${\chi}^F$,
then
\begin{equation}
	\vec{L}_S \phi = [{\vec{L}}^{(\rho)}{\psi}_S^m]
	{\chi}^F_m + {\psi}_S^m
	[{\vec{L}}^{(\overline{\rho})}{\chi}^F_m].  
\end{equation}
Hence
\begin{equation}
	(\vec{{\cal L}_S})_{mn} =
	{\delta}_{mn}{\vec{L}}^{(\rho)} + ({\chi}^F_m,
	[{\vec{L}}^{(\overline{\rho})}{\chi}^F_n]) 
\end{equation}
This then is the correct way to write $\vec{{\cal L}_S}$. We will
henceforth follow the correct path and abandon the erroneous (\ref{curlyL}). 

\subsection{On How Parity Doubles are Split}
There is nothing here that cannot be inferred from \cite{BohMot2}. But
although not new, what follows is helpful to understand how the degeneracy
of the doubles is lifted.  

In the state ${\chi}^F_m$, the kets $|m \rangle$ carry the UIR
$\overline{\rho}$ of $H^*$, while the group $SU(2)$ acts on $|k,j
\rangle$ with generators $\vec{L}_F$. The total angular momentum is thus  

\begin{equation}
	\vec{J}_T = \vec{L}_F + \vec{L}_S.
\end{equation}
The state ${\chi}^F_m$ is a singlet under $\vec{J}_T$ because 

\begin{eqnarray}
	e^{i \vec{\theta}.\vec{J}_T} \{|m',j \rangle
		D^j_{m'm}(g) \} &=& \{ U(e^{i \vec{\theta}.
		\vec{\sigma}/2}) |m',j \rangle \} \{ D^j_{m'm}(e^{-i
		\vec{\theta}. \vec{\sigma}/2}g) \}
		\nonumber   \\
	 &=& |m',j \rangle D^j_{m'm}(g).
\end{eqnarray}
Thus
\begin{equation}
	\vec{J}_T\phi = (\vec{L}_S^{(\rho)} \otimes
		{\bf 1})\phi := [\vec{L}_S^{(\rho)}{\psi}_S^m] {\chi}^F_m  .
\end{equation}
With this result in mind, we write 
\begin{eqnarray}
	\vec{L}_S &=&\vec{L}_S^{(\rho)} \otimes {\bf 1}+
		{\bf 1} \otimes \vec{L}_S^{(\overline{\rho})}
		\nonumber \\ 
		&=& \vec{L}_S^{(\rho)} \otimes {\bf 1} + {\bf 1}
		\otimes (\vec{J}_T-\vec{L}_F)
\end{eqnarray}
and find

\begin{equation}
	\vec{L}_S\phi = [\vec{J}_T {\psi}_S]^m
		{\chi}^F_m - {\psi}_S^m[\vec{L}_F{\chi}^F]_m \;. 
\end{equation}
In the leading B-O approximation, $\vec{L}_F$ gets restricted to the span
of ${\chi}^F_m$ for $m \in \overline{I}$ in the computation of $\vec{{\cal
L}_S}$. We will now see that our doubles are split by $\cal H$ on removing
this restriction.  

The states $\phi$ and ${\cal K} \phi$ are not generally orthogonal. There
is no reason for them to be so.  

Let us first concentrate on $\cal K$ being $\cal P$ and form the
orthogonal ${\cal P} = \pm 1$ states, the superposition in its
formation being permitted as both $\phi$ and ${\cal P} \phi \in
V^{({\rho}_0)}$:

\begin{eqnarray}
	|\pm \rangle & = \frac{\phi \pm {\cal P} \phi}{\sqrt{2}[1 \pm
		(\phi,{\cal P}\phi)]^{1/2}},  \\
	{\cal P}|\pm \rangle & = \pm |\pm \rangle.
\end{eqnarray}
Here we have used the hermiticity of $\cal P$ and the attendant reality of
$(\phi,{\cal P}\phi)$. For these states, 

\begin{eqnarray}
	{\cal H}|\pm \rangle &=& [\frac{1}{2{\cal I}} ({\vec{J}_T}^2
		+ \vec{L}_F^2) + {\cal H}_F]|\pm \rangle \nonumber \\ 
	    & &	- \frac{1}{{\cal I} 
		\sqrt{2}[1 \pm (\phi,{\cal P}\phi)]^{1/2}} \{
		(\vec{J}_T{\psi}^m_S) (\vec{L}_F{\chi}^F_m) \pm
		(\vec{J}_T{\cal P}{\psi}^m_S) (\vec{L}_F{\cal P}{\chi}^F_m) \}.
\label{Pestates} 
\end{eqnarray} 
We can diagonalise ${\vec{J}_T}^2$ along with $\cal H$ so that the first
term is proportional to $| \pm \rangle$. The ground state angular momenta
for $| \pm \rangle$ would also be equal. [The total angular momentum and
magnetic quantum number labels $J$ and $\sigma$ are for now suppressed.] So
the first term does not split the energies of $| \pm \rangle$. As for the
second term, as ${\vec{L}_F}^2$ has value $j(j+1)$ on $|m,j \rangle$, it
serves to correct the equation determining ${\psi}_j$. It cannot help to
split the parity doubles if ${\cal H}_F$ fails to do so.  

But the last term in (\ref{Pestates}) would generally split the doubles. Its
scalar products with $| \pm \rangle$, which are the mean interaction
energies from the perturbing Hamiltonian $-\vec{J}_T.\vec{L}_F /{\cal I}$,
are not the same for ${\cal P} = \pm 1$ because of the differing
normalization factors $[1 \pm (\phi,{\cal P}\phi)]^{1/2}$ of (\ref{Pestates}).

In this manner, we see that the parity doubles are not degenerate in the
exact theory. 

\subsection{Split $\cal P$- Doubles are $\cal T$- Doubles Too}
We will assume in further work that the doubles $|+ \rangle$ and
$|-\rangle$ are certainly split. Now ${\cal T}|\pm \rangle$ is degenerate
with $|\pm \rangle$ and hence by hypothesis cannot be $|\mp
\rangle$. Therefore $\cal T$ does not affect the eigenvalues of $\cal P$
and $\cal P$ and $\cal T$ commute: 

\begin{equation}
	{\cal P}{\cal T} = +{\cal T}{\cal P}.
\label{PTcommute}
\end{equation}

If $\phi$ has total angular momentum $J$ and its third component $\sigma$,
we can write 
\begin{equation}
	\phi = |\sigma,J \rangle \chi
\end{equation}
where $|\sigma,J \rangle$ are orthonormal vectors transforming by the
standard rotation matrices \cite{Edmonds,Rose,BLCM} and $\chi$ are
annihilated by $\vec{J_T}$. [We assume that there is no total
angular momentum degeneracy for fixed energy for the states in question.]
The unit norm of $\phi$ fixes the norm of $\chi$.

It follows that the even and odd parity states can be taken to be
\begin{eqnarray}
	|\pm;\sigma,J \rangle &=& \frac{1}{\sqrt{2}[1 \pm
		(\phi,{\cal P}\phi)]^{1/2}} ({\bf 1} \pm {\cal P})|\sigma,J
		\rangle \chi, \\
	{\cal P}|\pm ;\sigma,J \rangle &=& \pm |\pm ;\sigma,J \rangle 
\end{eqnarray}
where now the angular momentum labels of the states are also displayed. 

The state ${\cal T}|\epsilon;\sigma,J \rangle \quad [\epsilon = \pm
1]$ is linear in $|\epsilon;J,\sigma \rangle$ by (\ref{PTcommute}) and
transforms under $SU(2)$ according to 

\begin{equation}
	e^{i \vec{\theta}.\vec{J}_T}({\cal T}
		|\pm;\sigma,J \rangle )=({\cal T}|\pm;{\sigma}',J \rangle
		)D^J_{{\sigma}'{\sigma}}(e^{i \vec{\theta}.\vec{\sigma}/2})^* 
\end{equation}
because $e^{i \vec{\theta}.\vec{J}_T}$ commutes with $\cal T$ and 
\begin{equation}
	e^{i \vec{\theta}.\vec{J}_T}
		|\pm;\sigma,J \rangle =|\pm;{\sigma}',J \rangle
		D^J_{{\sigma}'{\sigma}}(e^{i \vec{\theta}.\vec{\sigma}/2}).
\end{equation}
It follows that 
\begin{equation}
	{\cal T}|\pm;\sigma,J \rangle = |\pm;{\sigma}',J \rangle
		C_{{\sigma}'{\sigma}} 
\end{equation}
where a possible phase on the right hand side has been set equal to 1
and $C$ implements the equivalence between $D^J$ and ${D^J}^*$, 

\begin{equation}
	C^{-1} D^J C = {D^J}^*,\quad C^2 = (-1)^{2J}{\bf 1}.
\end{equation}
$C$ is the reduced rotation matrix for $\pi$ rotation for the conventional
phase choices of angular momentum matrices wherein the second angular momentum
component is purely imaginary and the first and third components are real:
\begin{eqnarray}
	C &= D^J[e^{i\pi {\sigma}_2/2}], \label{Cdef1} \\
	C_{{\sigma}'{\sigma}} & = (-1)^{J - \sigma}
		{\delta}_{\sigma,-{\sigma}'}. \label{Cdef2}
\end{eqnarray}

Note that 
\begin{equation}
	{\cal T}^2 = ({\cal PT})^2 = (-1)^{2J}.
\end{equation}

Wigner has elsewhere shown \cite{gursey} that we would have these identities if
$\cal P$ and $\cal T$ are implemented on a space of states carrying a
single UIR of spin cover of the rotation group.  

\subsection{Split $\cal T$-Doubles with the Same Parity}
These occur for staggered conformations. Our assumption as before is
that the doubles are certainly split. Given this hypothesis, we want
to know the sort of linear combinations of states diagonalising $\cal
H$. 

Let us write
\begin{eqnarray}
	\phi &:= |\sigma,J \rangle_1 \chi \nonumber, \\
	|\sigma,J \rangle_1 & \equiv |\sigma,J \rangle
\end{eqnarray}
and also let
\begin{eqnarray}
	{\cal T} \phi &= {\cal T}(|\sigma,J \rangle_1 \chi) \nonumber \\
		      &= |\sigma',J \rangle_2 C_{\sigma' \sigma}\; \chi^* .
\label{Tphi}
\end{eqnarray}
Here $|\sigma',J \rangle_2$ serves the role of $|\sigma,J \rangle$ for
${\cal T}\phi$ and $C$ accounts for the fact that ${\cal T}\phi$
transforms by $D^{J^*}$ under $SU(2)$. As in the case of $\cal P$, there is
no reason for $|\sigma,J \rangle_2 \chi^*$ to be orthogonal to $|\sigma,J
\rangle_1 \chi$.

As for ${\cal T}({\cal T}\phi)$, we can write \cite{gursey} 

\begin{equation}
	{\cal T}^2 \phi=\eta_T \phi,\quad \eta_T= +1\;{\rm or}\;-1.
\label{T2phi}
\end{equation}
That being so, we have, 
\begin{equation}
	{\cal T}(|\sigma',J\rangle_2 \chi^*) = \eta_T |\sigma'',J \rangle_1
		C^{-1}_{\sigma'' \sigma} \; \chi,
\label{Tsigma}
\end{equation}
since $C^*=C$ under prevailing conventions [Cf (\ref{Cdef1}) and
(\ref{Cdef2})].  
States of definite energy carry an irreducible representation of
$\vec{J}_T$ and so have a basis of vectors  
\begin{equation}
	z_1 |\sigma,J \rangle_1\; \chi + z_2 |\sigma,J \rangle_2\; \chi^*,
		z_i \in {\bf C}.
\end{equation}

We want to determine $z_i$ to the extent possible by general arguments. Let
$W$ be the vector space spanned by these states for fixed $z_i$. These
states then have a fixed energy. If ${\cal T}W \cap W = \{ 0 \}$, then $W
\oplus {\cal T}W$ will consist of all the $\cal T$-doubles, and they would
be degenerate too since $w \in W$ and ${\cal T}w$ have the same energy. We
have excluded this possibility by assumption. Now both $W$ and ${\cal T}W$
carry a UIR of $SU(2)$ and so ${\cal T}W \cap W$ is either $W(={\cal T}W)$
or $\{ 0 \}$. Having already set aside the last possibility, we have

\begin{equation}
	{\cal T}W = W.
\end{equation}
Hence, remembering that $\cal T$ reverses $\sigma$ (and assuming that there
 is no angular momentum degeneracy for fixed energy for these states), 

\begin{eqnarray}
	{\cal T}(z_1 |\sigma,J \rangle_1\; \chi + z_2 |\sigma,J
		\rangle_2\; \chi^*) & = & \omega_{\sigma}[z_1|-\sigma, J
		\rangle_1\; \chi + z_2 |-\sigma,J \rangle_2 \;
		\chi^* ], \label{Tgen1} \\
	{\cal T}^2 (z_1 |\sigma,J \rangle_1\; \chi + z_2 |\sigma,J
		\rangle_2\; \chi^*) & = & \omega_{\sigma}^*
		\omega_{-\sigma} (z_1|\sigma,J \rangle_1\; \chi + z_2
		|\sigma,J \rangle_2\; \chi^*)   \label{T2gen1}
\end{eqnarray}
where
\begin{equation} 
|\omega_{\sigma}| = 1, \omega_{\sigma} \in {\bf C}.
\end{equation}
But from (\ref{Tphi}), (\ref{T2phi}) and (\ref{Tsigma}),

\begin{eqnarray}
	{\cal T}(z_1 |\sigma,J \rangle_1\; \chi + z_2 |\sigma,J
		\rangle_2\; \chi^*) & = & z_1^* |\sigma',J \rangle_2
		C_{\sigma' \sigma}\; \chi^* + \eta_T z_2^* |\sigma',J
		\rangle_1 C^{-1}_{\sigma' \sigma}\; \chi,   \label{Tgen2} \\
	{\cal T}^2 (z_1 |\sigma,J \rangle_1\; \chi + z_2 |\sigma,J
		\rangle_2\; \chi^*) & = & \eta_T (z_1 |\sigma,J
		\rangle_1\; \chi + z_2 |\sigma,J \rangle_2\; \chi^*).
		\label{T2gen2} 
\end{eqnarray}
Comparing (\ref{Tgen1}) and (\ref{T2gen1}) with (\ref{Tgen2}) and
(\ref{T2gen2}), we get 

\begin{eqnarray}
	\omega_{\sigma}^* \omega_{-\sigma} &=& \eta_T, \label{ww}\\
	z_1 \omega_{\sigma} &=& \eta_T z_2^* (-1)^{J + \sigma}, \label{z1w}\\
	z_2 \omega_{\sigma} &=& z_1^* (-1)^{J - \sigma} \label{z2w}.
\end{eqnarray}
Either of the last two equations gives 
\begin{equation}
|z_1| = |z_2| = {\rm a \;constant}\; \lambda.
\end{equation}
Hence,
\begin{equation}
	z_i \neq 0.
\end{equation}
Set 
\begin{equation}
	\omega_{\sigma} = (-1)^{J + \sigma} \omega.
\label{w}
\end{equation}
Then,

\begin{equation}
	|\omega_{\sigma}|=1 \Rightarrow |\omega| = 1.
\end{equation}
Now $\omega$ is independent of $\sigma$ by (\ref{z1w}) since $z_i \neq
0$. Hence by (\ref{ww}), (\ref{z1w}) and (\ref{z2w}),

\begin{eqnarray}
	\eta_T &=& (-1)^{2J}, \\
	z_1 \omega &=& \eta_T z_2^*
\end{eqnarray}
so that 

\begin{eqnarray}
	z_i &=& \lambda e^{i \theta_i} ,  \\
	\omega &=&\eta_T e^{-i (\theta_1 + \theta_2)},  \\
	\theta_i &=& {\rm real}
\end{eqnarray}
and

\begin{equation}
	z_1 |\sigma,J \rangle_1\; \chi + z_2 |\sigma,J \rangle_2\; \chi^* =
		\lambda \{ e^{i \theta_1} |\sigma,J \rangle_1 \chi
		+ e^{i \theta_2} |\sigma,J \rangle_2 \chi^* \}.
\end{equation}

In a particular model, states of a particular fixed energy will have a
fixed value $\tilde{\theta_i}$ for $\theta_i$ and will be spanned by
\begin{equation}
	e^{i \tilde{\theta_1}} |\sigma,J \rangle_1\; \chi
	+ e^{i \tilde{\theta_2}} |\sigma,J \rangle_2\; \chi^* .
	 \label{Tstates1}    
\end{equation}
Their $\omega$ will also be fixed to be some $\tilde{\omega}$:
\begin{equation}
	\omega = \tilde{\omega} = \eta_T e^{-i (\tilde{\theta_1} +
		\tilde{\theta_2})}.  
\end{equation}
Another manifold of states with a fixed energy split from (\ref{Tstates1})
will be spanned by vectors orthogonal to (\ref{Tstates1}). These two sets
of states span $W \oplus {\cal T}W$.  

There remains the freedom in certain phase choices. Thus let

\begin{eqnarray}
	\widetilde{|\sigma,J \rangle_i} &=& e^{i \tilde{\theta_i}}
		|\sigma,J \rangle_i,  \\
	\tilde{\cal T} &=& e^{i (\tilde{\theta_1} +
		\tilde{\theta_2})}{\cal T}.
\end{eqnarray}
Then,

\begin{eqnarray}
	\tilde{\cal T} \widetilde{|\sigma,J \rangle_1} &=&
		\widetilde{|\sigma',J \rangle_2} C_{\sigma' \sigma}, \\
	\tilde{\cal T} \widetilde{|\sigma,J \rangle_2} &=&
		\widetilde{|\sigma',J \rangle_1} C_{\sigma' \sigma}, \\
	{\tilde{\cal T}}^2 &=& (-1)^{2J},
\end{eqnarray}
and (\ref{Tstates1}) takes the simple form 

\begin{equation}
	\widetilde{|\sigma,J \rangle_1}\chi +
		\widetilde{|\sigma,J \rangle_2}\chi^*.
\end{equation}

\section{$\cal P-,T$- Doubles in Quark Physics}

There are several areas of particle physics with the potentiality to
support $\cal K$-doubles. As always, it can be realized only when
there are two well-separated time scales $T_S$ and $T_F$,

\begin{equation}
	T_S/T_F \gg 1.
\label{SF}
\end{equation}
In addition, the expected life time $\tau$ of a suspected double must
be large compared to $T_S$:

\begin{equation}
	\tau /T_S \gg 1
\label{tauS}
\end{equation}
If this inequality is violated, and a state decays before its slow core
completes several revolutions, even unstable doubles are not likely to
occur. 

Let us now briefly examine typical multi-quark states which may
support these doubles. Other areas of particle physics where the B-O
approximation may work will be discussed in Section 8.

\subsection{Quark Physics}
In a previous paper \cite{bava}, we had examined three-quark
systems, two of them forming the heavy core, and checked if there are
favorable candidates compatible with (\ref{SF}) and (\ref{tauS}). Our
conclusion was that the spectra of $cbu$ and $cbd$ baryons are the best
places to look for $\cal P-,T$- doubles. The estimates for (\ref{SF}) and
(\ref{tauS}) were 
\begin{eqnarray}
	T_S/T_F &\sim& 8.8-11, \\
	\tau /T_S &\sim& {10}^9.
\end{eqnarray}
We will argue below that these B-O states are not quark model states. They
are also not covered by models using heavy quark symmetry (see
\cite{neubert} and references therein for a recent review of heavy quark
symmetry) which typically have just one heavy quark.  

In course of time, states with many quarks having properties like
molecules would surely be found. Dibaryons such as the $H$ with six
quarks have already been predicted \cite{jaffe1,jaffe2,bblrs,blrs} and
searched for \cite{mas,ad1,ad2}. Six-quark states for example, with a heavy
core populated by $c$'s and $b$'s and a light cloud of $u$'s and $d$'s
would be ideal for the successful application of the B-O approximation and
hence also the search for $\cal P-,T$- doubles. But the experimental
formation of such states are obviously extremely hard for now.  

Other possibilities of this sort would be states with both quarks and
antiquarks. For example, we can look at $(c \overline{b})(u
\overline{d})$. This is really the composite state of a heavy meson $c 
\overline{b}$ with a light one $u \overline{d}$ and the B-O
approximation should work if $c \overline{b}$ can be described by an
arrow. But further study would be needed if a more complicated
description along the lines of the chiral Lagrangian is called for. 

\subsection{Born-Oppenheimer States Are Not Quark Model States}
In the quark model for mesons and baryons, it is generally the case
that the total orbital angular momentum of the quarks has a fixed
value. We will now prove that the contrary is correct for the B-O wave
function showing that these two models are different. 

Let us write the fast wave function in the form (\ref{fastwf}). It has zero
total angular momentum. So if $\phi$ has total angular momentum $J$ and
magnetic quantum number $\sigma$, ${\psi}_S^m$ and $\phi$ must have the forms 

\begin{eqnarray}
	{\psi}_S^m(g) &=& c_J D^J_{\sigma m}(g),\quad m \in \overline{I}, \\
	\phi(g,.) &=& c_J {\sum}_{m,j}D^J_{\sigma m}(g)|k,j \rangle
		D^j_{km}(g){\alpha}_j(.) \label{qstate}
\end{eqnarray}
where $c_J$ is to be tuned to get 1 for the norm of $\phi$ and where, as
per (\ref{3.8}), the index set $\overline{I}$ is such that 

\begin{equation}
	D^j_{km}(gh)= \sum_{m' \in \overline{I}} D^j_{km'}(g)
		\overline{\rho}_{m'm}(h).
\end{equation}

Orbital rotations are spatial rotations which do not affect spin and
hence the kets $|k,j \rangle$. Therefore they are the $SU(2)$
transformations on the left of $g$. Orbital angular momentum is then
not sharp on the state (\ref{qstate}), the latter being a superposition of
states of orbital angular momenta from $|J-j|$ to $J+j$. We thus see
that the quark model and the B-O approximations are different. 

\subsection{Signals For $\cal P-,T$- Doubles}
In molecular physics, there is a neat way of experimentally detecting
parity doubles. It goes as follows \cite{Herzberg2}.

Low energy excitations of molecules are rotational bands stacked on
vibrational energies $E_{n}$ (see for example, \cite{LL-QMNR}). For a
molecule with moment of inertia ${\cal I}$, they have energies $E_{n} +
J(J+1)/2{\cal I}$ with the angular momentum $J$ assuming successive
values. The separation $E_{n'} - E_{n}$ of vibrational excitations is much
larger than rotational energies. Now if the levels $(n, J)$ for given $n$
and $J$ are non-degenerate (but for angular momentum degeneracy), then one
of the transitions $(n', J) \rightarrow (n, J)$ or $(n', J \pm 1)
\rightarrow (n, J)$ would be forbidden in the dipole approximation by
parity conservation, and the corresponding spectral line would be
weak. This is so because in this scenario, states of successive $J$ and
same $n$ differ in parity. In this way, one can indirectly infer the
existence of parity doubles. 

As for $\cal T$- doubles with the same parity, we have not found any
discussion of their experimental detection in chemistry. These doubles are
similar to the ones that occur in the presence of Kramers'
degeneracy (see for example, \cite{Merzbacher}). Their existence is usually
inferred indirectly, from statistical properties, as in the case of
diamagnetic susceptibility of certain rare earth elements.   

In nuclear physics, there is no direct experimental detection of $\cal
P$-doubles and there is no reported example of a $\cal T$- double of the 
same parity. Theory, involving rather elaborate calculations, predicts
$\cal P$-doubles at certain energies,and when experiments find
excitations with these energies, they are accepted as the predicted
doubles \cite{lesh,lech}. 

We can think of no clean signals for the detection of $\cal P$- or
$\cal T$- doubles in quark physics either. How is one to
experimentally tell apart a B-O state from a quark model state?
Lacking unambiguous means for this purpose, theory should be the
ultimate judge in this matter for now.

\section{Towards a general B-O theory}
In previous sections, we always thought of $Q$ as $SU(2)/H^*$ and
assumed rotational invariance. While these assumptions are good for
illustrative purposes, they are also limiting when $Q$ is a general
manifold and the system lacks an obvious and appropriate
symmetry. Such being often the case in physics, it is progressive to
work towards a general theory of the B-O approximation. That is what
we will try here in this section. There are certainly many works of B-O
approximation which are not dependent on symmetry. The novelty here is its
emphasis on bundle theory and topology. 

In the B-O approach, we are not concerned with the whole Hilbert space
of the fast wave functions. That would be useless as any two
infinite-dimensional (separable) Hilbert spaces are isomorphic. Rather we
focus on the space of eigenstates $\chi^F$ of the fast Hamiltonian ${\cal
H}_F$ for a fixed energy in the discrete spectrum. Often, this eigenspace
corresponds to the set of ground states of ${\cal H}_F$. Being eigenstates
of ${\cal H}_F$, they are also in the domain of ${\cal H}_F$ and are thus
associated with smooth sections of a vector bundle over the fast
configuration space $\{ F \}$. But this bundle structure is not central in
the B-O approach and will not be mentioned henceforth. 

That is not all: there is more to be said about fast dynamics. In the
B-O approximation, we solve for $\chi^F$ imagining that the slow variables
are static. So $\chi^F$ depends not just on $\{ F \}$, but is sensitive
also to $q \in Q$ as a sort of background variable. 

But there is no particular reason for $\chi^F$ to be a suitably smooth
function on $Q$. It is quite enough if the scalar product $(\chi^F,
\widetilde{\chi^F})$ of any two eigenstates of ${\cal H}_F$ for the
energy of interest is a smooth function on $Q$. The reason is that only
probability densities are observable and wave functions are not. 

But while $\chi^F$'s may not be smooth functions of $Q$, it turns out
that they are multi-valued functions on $Q$. Their nature can be
understood in the following manner modelled  on the work of Shapere and
Wilczek reproduced in \cite{ShaWil}.

Let us first fix a fiducial point $q_0 \in Q$. Assume that the slow
system is initially at $q_0$ and that we have a corresponding fast
wave function $\chi^F(q_0,.)$. Imagine now that the  slow system is 
adiabatically dragged around in a loop $\Gamma_{q_0}$ ending up again
at $q_0$. That is, we slowly transport the slow system from $q_0$ to
$q_0$ along this loop. All along this process, the fast wave function
remains an eigenstate with energy varying continuously. Being in the
point spectrum, there is no uncertainty about its value.[Level
crossings are assumed not to occur]. So after the circuit
$\Gamma_{q_0}$, energy returns to its original value. But what about
$\chi^F$? 

$\chi^F$ need not return to its original value $\chi^F(q_0,.)$ but can
undergo a unitary transformation. If $N$ is the dimension of the
eigenspace and $\chi^F_m(q_0,.)$ the components of $\chi^F$ in some
orthonormal basis for the eigenspace at $q_0$, then it could happen
that 

\begin{equation}
	\chi^F_m(q_0,.) \rightarrow
		\chi^F_{m'}(q_0,.)u(\Gamma_{q_0})_{m'm},
\end{equation}
$u(\Gamma_{q_0})$ being an $N \times N$ unitary matrix. This 
transformed wave function has the same eigenvalue as $\chi^F_m(q_0,.)$
and is normalized, so such a change in $\chi^F_m(q_0,.)$ can
happen. 

Let $\Gamma_{q}$ be an unparameterized path from $q_0$ to
$q$ so that only its geographical location in $Q$ matters, and let
$\wp = \{ \Gamma_{q} \}$ the path space of $Q$ with base point $q_0$
\cite{zsnmmb,BMSS}. [The loop $\Gamma_{q_0}$ above is a member of $\wp$.]
It is clear now that $\chi^F$ is not best thought of as a function on $Q$,
but we can think of them as special sorts of functions on $\wp$. Writing
$\chi^F(\Gamma_{q},.)$ for the wave function $\chi^F(q,.)$ at $q \in Q$, their 
specialty comes from the property  

\begin{equation}
	\chi^F_m(\Gamma_{q_0} \cup \Gamma_{q},.) =
		\chi^F_{m'}(\Gamma_{q},.)u_{m'm}(\Gamma_{q_0}). 
\label{pathstate}
\end{equation}
The curve $\Gamma_{q_0} \cup \Gamma_{q}$ here is obtained by concatenation:
one first travels $\Gamma_{q_0}$ and then $\Gamma_{q}$. 

The space $\wp$ is a fibre bundle over $Q$. The projection map is 
\begin{eqnarray}
	\pi : \wp & \rightarrow Q  \\
		\Gamma_{q} & \rightarrow q
\end{eqnarray}
and the fibre consists of all paths ending up at $q$.

The transformation law (\ref{pathstate}) is reminiscent of sections of vector
bundles \cite{abbjrs2}. But $\Gamma_{q_0}$'s do not form a group when
they are composed by concatenation of curves. We need to impose more
structure on $\Gamma_{q}$ to get a vector bundle out, and we shall
soon do so. 

As an eigenstate of ${\cal H}_S$ or a wave function in its domain need
not also be a function on $Q$, but can always be thought of as a
function on $\Gamma_{q}$ with an equivariance property like
(\ref{pathstate}). That is enough to ensure that probability densities
therefrom are functions on $Q$. We had previously made the trivializing
assumption that equivariance was in fact invariance so that this wave
function was a function on $Q$. Then as we saw, the slow wave function
$\psi_S$ has a transformation law (\ref{slowwftrn}) with $u$ replaced by its
complex conjugate $u^*$ while the total wave function $=(\psi_S^m
\chi_m^F)$ is a function on $Q$. 

When $N > 1$, there is a redundancy in aspects of the description
wherein we allow both components of $ \chi^F$ and the argument
$\Gamma_{q}$ to vary for a fixed $q$ \cite{BMSS}. But there is no
need for us to be distracted by this issue: it is only of minor
relevance to the present work, and does not damage the conclusions. 

We shall now reproduce known examples of $\chi^F$ by postulating
particular dependences of $u$ on $\Gamma_{q_0}$. They will turn $u$'s
into representations of groups and thereby also lead us to the sought-for
vector bundles.  

\subsection{Flat Bundles}
The nature of $\chi^F$ here is governed by the fundamental group
$\pi_1(Q)$ of the manifold $Q$ \cite{BMSS}. Let $q_0$ be the base point
for defining the homotopy groups and let us assume that $u$ depends
only on the homotopy class $\langle \Gamma_{q_0} \rangle$ of the loop
$\Gamma_{q_0}$. We can then write $u(\Gamma_{q_0})$ as $u(\langle
\Gamma_{q_0} \rangle)$. It is also easy to see that $u$ defines a
unitary representation of $\pi_1(Q)$. Let us assume it to be
irreducible, the general case being a direct sum of UIR's. This case
then corresponds to $\chi^F$ being sections of associated bundles for
a UIR of $\pi_1(Q)$ \cite{abbjrs2}, a result which can be briefly
explained as follows. 

Let $\langle \Gamma_{q} \rangle$ be the equivalence class of paths
homotopic to $\Gamma_{q}$. An element of the vector bundle for the
representation $u$ is an equivalence class 

\begin{equation}
	(\langle \Gamma_{q_0} \rangle,v),\quad v=(v^1, v^2,...v^N),
\end{equation}
the equivalence relation being 

\begin{equation}
	(\langle \Gamma_{q_0} \rangle,v)=(\langle \Gamma_{q_0} \cup
		\Gamma_{q} \rangle ,vu(\langle \Gamma_{q_0} \rangle)).
\end{equation}
The section defined by $\chi^F$ is just 

\begin{equation}
	q \rightarrow (\langle \Gamma_{q} \rangle, \chi^F(\langle
		\Gamma_{q} \rangle ,.)).
\end{equation}
So $v$ above should be a function of $\{ F \}$ [just as $\chi^F(\langle
\Gamma_{q} \rangle ,.)$] to account for our case, but that is only a
modest change. 

In the same manner, $\psi_S$ too can be regarded as a function of
$\langle \Gamma_{q} \rangle$ and defines a section of a vector
bundle. Let us assume for illustration that ${\cal H}_S$ acts on functions
on $Q$. Once that is so, we can verify as before that differentials $d$ on
$Q$ which would occur in ${\cal H}_S$ become covariant differentials
$\nabla$ in the B-O approximation with the transformation 

\begin{equation}
	\nabla \rightarrow u[\langle \Gamma_{q} \rangle]^{-1} \nabla
		u[\langle \Gamma_{q_0} \rangle]
\end{equation}
when $\chi^F(\langle \Gamma_{q} \rangle, .)$ is acted on by
$\pi_1(Q)$:

\begin{equation}
	\chi^F(\langle \Gamma_{q} \rangle ,.) \rightarrow
		\chi^F(\langle \Gamma_{q_0} \cup \Gamma_{q} \rangle
		,.)  
\end{equation}
So if ${\cal H}_S$ acts on functions on $Q$, the relevant UIR for
$\psi_S$ is $u^*$. 

It remains to write the total wave function $\phi$. The best way to
write it is probably 

\begin{equation}
	\phi(\langle \Gamma_{q} \rangle ,.)=\sum_{\langle \Gamma_{q_0}
		\rangle \in \pi_1(Q)} \psi^m_S(\langle \Gamma_{q_0}
		\cup \Gamma_{q} \rangle) \chi^F_m(\langle \Gamma_{q_0}
		\cup \Gamma_{q} \rangle, .) .
\label{totalwf}
\end{equation}
Being $\pi_1(Q)$-invariant, this $\phi$ would give us a section of a
trivial bundle over $Q$ as we want. 

The omission of the $m$-sum in (\ref{totalwf}) is intentional. Terms with
different $m$ are equal so that $m$ can be frozen to any fixed value. We 
can prove this by writing (\ref{totalwf})  as

\begin{equation}
	\sum_{\langle \Gamma_{q_0} \rangle \in \pi_1(Q)}
		\psi^S_{m'}(\langle \Gamma_{q} \rangle)
		u^*_{m'm}(\langle \Gamma_{q_0} \rangle)
		\chi^{m''}_F(\langle \Gamma_{q} \rangle)
		u_{m''m}(\langle \Gamma_{q_0} \rangle).  
\end{equation}
An easy application of orthogonality relations between $u$'s \cite{BalTra}
gives 
\begin{equation}
	\sum_{\langle \Gamma_{q_0} \rangle \in \pi_1(Q)}
		u^*_{m'm}(\langle \Gamma_{q_0} \rangle)   
		u_{m''m}(\langle \Gamma_{q_0} \rangle)= \delta_{m'm''}/N
\end{equation} 
which shows the result.

Elsewhere \cite{zsnmmb,BMSS}, we have explained that $\{\langle \Gamma_{q}
\rangle \}$ is just the universal cover $\tilde{Q}$ of $Q$.

\subsection{$U(1)$ Bundles}
Let us go step-by-step and see how to get $U(1)$ bundles next. They
are important for a wide variety of physical systems including the
lowly $N_2O$.

The wave function $\chi^F$ can always be thought of as a function on
$\wp$. By  the general argument above, we have the transformation law
(\ref{pathstate}). If for a particular class of $\Gamma_{q_0}$, it happens
that $u(\Gamma_{q_0}) = 1$, we can then identify $\Gamma_{q}$ and
$\Gamma_{q_0} \cup \Gamma_{q}$ thereby getting a smaller space on
which $\chi^F$ is well-defined. 

The universal cover $\tilde{Q}$ is just one such space. In that case,
$u$ depends only on $\langle \Gamma_{q} \rangle$. Assuming that $u$ is
a faithful representation of $\pi_1(Q)$, we can then say that 

\begin{equation}
	\tilde{Q} = \wp/\sim .
\end{equation}
where the equivalence relation is defined by

\begin{equation}
	\Gamma_{q} \sim \Gamma'_{q} \Leftrightarrow u(\Gamma_{q}) =
		u(\Gamma'_{q}) 
\label{equivalence}
\end{equation}

Occasionally, it is convenient to think of $u(\Gamma_{q})$ as the
Wilson integral of a flat connection on $\Gamma_{q}$ for this case:

\begin{equation}
	u(\Gamma_{q})=Pe^{\int_{\Gamma_{q}}A},\quad A = {\rm a \;flat
		\; connection} 
\label{uA}
\end{equation}

The circumstance under which $\chi^F$ becomes associated with a $U(1)$
bundle is similar and has also been explained before \cite{zsnmmb}. There
it is shown that the equivalence relation (\ref{equivalence}) turns
$\wp/\sim$ into a $U(1)$ bundle if $u$ has the form (\ref{uA}) where $A$ is
the connection of such a bundle. 

Under our assumptions where ${\cal H}_S$ acts on sections of a trivial
bundle, $\phi$ is associated with a trivial bundle as we have already
seen.  

\subsection{$G$-Bundles}
When $u$ is characterized by a flat bundle, we get $\tilde Q$, and
when it is given by a $U(1)$-connection, we get a $U(1)$ bundle. In
the same way, when $u$ has the expression (\ref{uA}) where $A$ is the
connection for a $G$-bundle on $Q$, $\wp/ \sim$ becomes the $G$-bundle as
explained in \cite{zsnmmb,bagoso}. In this case too, there is nothing
further to add about the nature of ${\psi}_S^m$ or $\phi$. 

\subsection{Level Crossings}

There is a serious deficiency in the above discussion. The same $Q$
can often support twisted bundles and connections of different
sorts, and we have not found any rule to tell which bundle and
connection will occur when. 

When $H^*$ is discrete, so that $Q=SU(2)/H^*$ has dimension three,
there is a general theorem saying that all bundles on $Q$ admit flat
connections \cite{BottTu}. In other words, wave functions are sections of
some associated $\pi_1(Q)$ bundles [ or rather of associated $U_N$
bundles reducible to these flat bundles]. But we do not know exactly
what the bundle and connection are without detailed knowledge of
$\chi^F$. The UIR $\overline{\rho}$ of $H^*$ on the fast eigenspace
does in fact fix the bundle, but it would be useful to know if the
connection $\nabla$ is flat or not. 

For molecules, $Q$ can be two-dimensional: it can be $S^2$ or
${\bf RP}^2$. For the former, we can in fact tell the bundle and the
connection with precision. If $K$ is the eigenvalue of $\vec{n}.\vec{L}_F$,
the former is the $U(1)$ bundle with Chern character $K$. Also, spatial
rotations can be lifted to the bundle. The connection compatible with this
lift is unique and is the well-known charge-monopole connection for the
same $K$, $4 \pi K$ being the product $eg$ of electric and magnetic charges
$e$ and $g$ \cite{BMSS}. So for $Q=S^2$, we can tell the nature of the
bundle by simple considerations. 

Let us also check out ${\bf RP}^2$. The latter admits the flat ${\bf Z}_2$
bundle $S^2$ and an infinity of rank 2 $D_{\infty}$-bundles as we saw in
Section 2. We can tell what the bundle is once more from the UIR of
$H^*$, but the properties of the connection remain vague without
further input. 

In one approach to the Berry phase \cite{ShaWil}, twisted bundles for fast
variables arise from level crossings. In this point of view, $Q$ can
be embedded in a larger space $\hat{Q}$. Also there are points in $\hat{Q}$
where two fast levels become degenerate, while they are not so in some
neighbourhood of these points. In particular they do not cross on $Q$. The
nature of the bundle on $Q$ and its connection are then inferred using
certain considerations developed by von Neumann and Wigner \cite{nw}. 

General discussions of this sort, at least for molecules, with good
candidates for $\hat{Q}$ would be worthwhile. We have not seen them in
the literature. 
 
\section{Final Remarks}
\subsection{Skyrmions and Monopoles}
Parity and time-reversal doubles may well occur in Skyrmion and grand
unified monopole physics \cite{bava}. For both Skyrmions and monopoles,
there now exist elaborate simulations of static configurations for
differing values of baryon number and monopole charge
\cite{brtoca,carson,himamu,suho1,suho2}. They are found as regular solids
with discrete symmetry groups. We can imagine that further calculations
will show static configurations such as a pear, with a $U(1)$ symmetry
group. Excitations with spin, like a $\rho$ or an $\omega$, or even a
nucleon, which can have non-zero helicity, can then lead to $\cal P$- and
$\cal T$- doubles for Skyrmions. As for monopoles, by attaching fast
constituents like a spin-half quark, we can hope to create these doubles in
monopole physics too.  

There is one potentially attractive application of the B-O method which is
not directly tied up with $\cal P,T$- doubles. That is the following: In the
original Skyrme approach \cite{BMSS,bnrs1,bnrs2,witten1,witten2}, the
soliton of the chiral model is interpreted as the nucleon. Spinorial
quantization of the soliton is possible because the configuration space $Q$
has double connectivity and the generator $\pi_1(Q)$ in the baryon or
winding number \cite{bnrs1,bnrs2} $\pm1$ sector is got by rotating the
soliton by $2 \pi$ [Cf \cite{BMSS} and references therein].  

There is a second description of the nucleon which couples quarks as
well to the chiral field. (See for example \cite{Bhaduri} and references
therein). Controversies concerning double-counting has been generated by
this approach.  

But there is a possible reconciliation of these two models. Consider
the chiral field coupled to three quarks and let us treat the former
as slow and the latter as fast. Assume furthermore that the slow
Hamiltonian ${\cal H}_S$ for the winding number 1 or -1 is to be
quantized with tensorial states. Our general considerations show that
the effective Hamiltonian $\hat{\cal H}_S$ after integrating out
quarks acts on spinorial states. Thus $\hat{\cal H}_S$ is the
Hamiltonian in the Skyrme approach whereas ${\cal H}_S+{\cal H}_F$ is
the Hamiltonian close to the model with explicit quark fields. 

There are details to be attended to before we can be confident of this
reconciliation. Work on these matters is in progress. 

\subsection{Heavy Meson Bound States}
Baryons, chiral solitons and monopoles are not the only favorable
systems for the application of the B-O ideas, even in particle
physics. Literature abounds in speculation suggesting the existence of
heavy meson bound states which can involve distinct mesons too
\cite{tor,mawi}. They can be the slow variables and suitable
excitations the fast ones and B-O approximation may be
applicable. There may even be $\cal P,T$ doubles among these mesons.

\subsection{Anomalies}

The derivation of anomalies using the adiabatic approximation is well
understood in field theory
\cite{fujikawa1,fujikawa2,faddeev,segal,kieu,federbush}. Typically, in this 
approach to anomalies, gauge fields are regarded as classical background
fields and anomalies are deduced from the response of the fermion
determinant or of its Fock vacuum to gauge transformations. We must think
of gauge fields as slow variables in this method and the spinorial fields as
fast. 

We found a similar situation when studying $\cal P,T$- doubles. When
there are $\cal P$- doubles for example, the ground state (say) of
${\cal H}_F$ is not $\cal P$-invariant just as the Fock vacuum is not
gauge invariant if there are gauge anomalies \cite{segal}. There is also
no sense in superposing the ground state $\chi^F$ and its $\cal P$-
transform ${\cal P}\chi^F$ and enforcing $\cal P$ invariance (or ${\cal 
P}=-1$), as the scalar product $(\chi^F, {\cal P}\chi^F)$ and the probability
densities of these superpositions are not functions on $Q$. We cannot
thus impose the condition ${\cal P}=1$ on a fast state, just as we
cannot impose gauge invariance on Fock vacuum if there is a gauge anomaly. 
 
But the situation changes for parity on including slow variables. The
derivatives in the slow Hamiltonian become covariant derivatives after
averaging over $\chi^F$. This twists the slow bundle inducing a $\cal
P$-anomaly there as well in such a way that there is no $\cal
P$-anomaly in the total slow times fast wave functions. We can form
total wave functions with ${\cal P}= \pm 1$ and if appropriate retain
only the ${\cal P}= +1$ state. 

It is natural to enquire if similar circumstances prevail in anomalous
gauge theories. If that were so, the anomaly in the gauge response of
the Fock vacuum would be cancelled by another anomalous response from
the gauge field wave function, and there would be no anomaly left in
the total state of the gauge and Fermi fields. [We remark here that
Federbush \cite{federbush} has found a regularisation of two-dimensional
chiral gauge theories with no axial anomalies. See also \cite{kieu} and
references therein.]

There are indications that this is exactly what happens. When the Fock
vacuum as constructed by Segal \cite{segal} is averaged out, the
derivative $\delta /{\delta A_i}$ in the Yang-Mills part acquires a
connection showing the anomalous gauge response of the gauge field
state. Just as for molecules, it can cancel the Fock space anomaly. The
calculations showing these results are very similar to those in molecular
physics.  

It seems that the effect of the connection in the slow system can be
accounted for by a gauged Wess-Zumino action in the Yang-Mills
Lagrangian. This modified Lagrangian $\hat{L_S}$ must be the
Lagrangian leading to the Hamiltonian $\hat{\cal H}_F$. 

A solution of this kind to the anomaly problem has been proposed
before by Faddeev \cite{faddeev}. We suggest that it can be justified in
the B-O approach as outlined above. 

These remarks on anomalies are only in the nature of a report on work
in progress. We hope to give a detailed account soon.  

\subsection{Quantum Gravity}
If $M_1$ and $M_2$ are two $N$-manifolds, their connected sum $M_1 \#
M_2$ is a new $N$-manifold. It is obtained by first removing $N$-balls
$B_N^{(i)}$ from $M_i$ and then identifying the boundary spheres of
$M_i \backslash B_N^{(i)}$. [For more details, see for example  \cite{BMSS}.]

In two and three dimensions, there exist the so-called prime manifolds
$P_i$ \cite{BMSS}. Any asymptotically flat two- or three-manifold with one
asymptotic region is the
connected sum of $R^2$ or $R^3$ and finitely many primes. The
two-dimensional primes are $T^2$ and $RP^2$, while there are infinitely
many primes in dimension three. 

Some time ago, Friedman and Sorkin \cite{frso1,frso2} suggested the
possibility of spatial slices in gravity theories with primes attached 
and argued that a prime is to be associated with an elementary
excitation, a quantum ``geon''. Friedman, Sorkin \cite{frso1,frso2}, 
Witt \cite{frwi,witt} and Surya \cite{sosu} have established that
${\pi}_1(Q)$ of the configuration space $Q$ in the presence of primes
is complicated, and implies, just as for Skyrmions or molecules, that
quantization for gravity is not unique. Quantum gravity with spinorial
geons may exist even for Einstein Lagrangian with no matter fields
\cite{frso1,frso2}. Quantum geons violating the familiar
spin-statistics connection can also be found  \cite{erice86,abbjrs2} in
canonical gravity.

We would like to suggest that even in conservative gravity which
ignores these quantization ambiguities and sticks to the trivial UIR of
${\pi}_1(Q)$, many other UIR's can turn up in the presence of fast
degrees of freedom. Thus for example a spinorial constituent attached
to a tensorial geon can, after integrating out the fast variables,
lead to an effective geon Hamiltonian $\hat{\cal H}_S$ for a spinorial
geon. In this way, we may induce many UIR's of ${\pi}_1(Q)$. We are
currently looking at this possibility. If correct, it would mean that
non-trivial UIR's of ${\pi}_1(Q)$ cannot be ignored even in
conservative gravity. 
\bigskip

\noindent {\bf Acknowledgments}

We have discussed the contents of this article with several
physicists. Special thanks are due to B. Dolan, M. V. N. Murthy, R. Sorkin and 
our experimental colleagues at Syracuse University for sharing their 
knowledge and insights with us. 

This work was supported in part by the US DOE under contract number
DE-FG02-85ER40231.  

\bibliography{parity}
\bibliographystyle{prsty}
\end{document}